\documentclass[oneside]{book}
\usepackage[utf8]{inputenc}
\usepackage{times}
\usepackage{geometry}
\usepackage{enumitem,amssymb}
\usepackage{ragged2e}
\usepackage{amsmath}
\usepackage{graphicx}
\usepackage{comment}
\usepackage{eurosym}
\usepackage{multicol}
\usepackage[usenames]{xcolor} %used for font color              
\definecolor{xlinkcolor}{cmyk}{1,1,0,0}
\usepackage{url}
\usepackage[
 colorlinks=true,    % false: boxed links; true: colored links 
 linkcolor=xlinkcolor,     % color of internal links            
 citecolor=xlinkcolor,     % color of links to bibliography
 filecolor=xlinkcolor,  % color of file links 
 urlcolor=xlinkcolor,      % color of external link
 final=true
]{hyperref}
\usepackage[super,sort&compress]{natbib}
\usepackage{enumitem}
\setenumerate{itemsep=0mm}
\usepackage{authblk}
\usepackage[belowskip=-5pt,aboveskip=-5pt]{caption}
\usepackage{subfiles}
\usepackage{tocbibind}
\usepackage[symbol]{footmisc}

\usepackage{color, colortbl}

\title{\vspace{-2cm}
$|\Delta \mathcal{B}| =2$: A State of the Field, and Looking Forward\\
\large{A brief status report of theoretical and experimental physics opportunities from the participants of\\
The Amherst Center for Fundamental Interactions Workshop\\
\textit{Theoretical Innovations for Future Experiments Regarding Baryon Number Violation} \\ %, Part 1}\\
In Coordination with the Snowmass 2021 Rare Processes and Precision Frontier}
}
\author[1]{K.S. Babu}
\affil[1]{Oklahoma State University}
\author[2,3]{Joshua Barrow\thanks{jbarrow3@vols.utk.edu; also at Fermilab}}
\affil[2]{The University of Tennessee at Knoxville}
\affil[3]{Fermi National Accelerator Laboratory}
\author[4,5]{Zurab Berezhiani}
\affil[4]{University of L'Aquila}
\affil[5]{Gran Sasso National Laboratory}
\author[6]{Leah Broussard\thanks{broussardlj@ornl.gov}}
\affil[6]{Oak Ridge National Laboratory}
\author[6]{Marcel Demarteau}
\author[7]{P.S. Bhupal Dev}
\affil[7]{Washington University, St. Louis}
\author[8,9]{Jordy de Vries\thanks{jdevries@umass.edu}}
\affil[8]{University of Massachusetts, Amherst}%, MA 01003, USA}
\affil[9]{RIKEN BNL Research Center, Brookhaven National Laboratory}%, Upton, NY 11973-5000, USA}
\author[10]{Alexey Fomin}
\affil[10]{NRC ``Kurchatov Institute" - PNPI}
\author[11]{Susan Gardner}
\affil[11]{University of Kentucky}
\author[12]{Sudhakantha Girmohanta}
\affil[12]{Stony Brook University}
\author[13]{Elena Golubeva}
\affil[13]{Institute for Nuclear Research, Moscow}
\author[14]{Maury Goodman}
\affil[14]{Argonne National Laboratory}
\author[15]{Julian Heeck}
\affil[15]{University of Virginia}
\author[16]{Yeon-jae Jwa}
\affil[16]{Columbia University} 
\author[2]{Yuri Kamyshkov}
\author[16]{Georgia Karagiorgi}
\author[17]{Bingwei Long}
\affil[17]{Sichuan University}
\author[18]{David McKeen}
\affil[18]{TRIUMF}
\author[19]{Prajwal Mohan Murthy}
\affil[19]{University of Chicago}
\author[20]{Rukmani Mohanta}
\affil[20]{University of Hyderabad}
\author[21]{Rabindra N. Mohapatra}
\affil[21]{University of Maryland}
\author[22]{Jean-Marc Richard}
\affil[22]{Universit\'e de Lyon \& IN2P3}
\author[23]{Enrico Rinaldi}
\affil[23]{Arithmer Inc. \& RIKEN iTHEMS}
\author[24]{Valentina Santoro}
\affil[24]{The European Spallation Source}
\author[12]{Robert Shrock}
\author[25]{W. M. Snow}
\affil[25]{Indiana University}
\author[26]{Anca Tureanu}
\affil[26]{University of Helsinki}
\author[3]{Michael Wagman\thanks{mwagman@fnal.gov}}
\author[27]{Linyan Wan}
\affil[27]{Boston University}
\author[28]{James D. Wells}
\affil[28]{University of Michigan, Ann Arbor}
\author[29]{Sze Chun Yiu}
\affil[29]{ Stockholms Universitet}
\author[30]{A. R. Young}
\affil[30]{North Carolina State University}
\date{}
\begin{document}

\maketitle

\frontmatter

%fix section numbering
\newcommand{\mychapter}[2]{
    \setcounter{chapter}{#1}
    \setcounter{section}{0}
    \chapter*{#2}
    \addcontentsline{toc}{chapter}{#2}
}

%section authors
\makeatletter
\newcommand{\chapterauthor}[1]{%
  {\parindent0pt\vspace*{-7pt}%
  \linespread{1.1}\normalsize\centering\scshape#1%
  \par\nobreak\vspace*{5pt}}
  \@afterheading%
}
\makeatother

\mychapter{0}{Foreword}%Introduction and Overview of the Workshop}

\iffalse
\begin{center}
    \href{https://indico.fnal.gov/event/44472/}{Theoretical Innovations for Future Experiments Regarding Baryon Number Violation} \\ %, Part 1}\\
%    \textit{Formerly \href{www.physics.umass.edu/acfi/seminars-and-workshops/prospects-for-baryon-number-violation-by-two-units}{Prospects for Baryon Number Violation by Two Units}}\\
    \textit{An \href{www.physics.umass.edu/acfi/}{Amherst Center for Fundamental Interactions Workshop}}\\
    In coordination with the Snowmass 2021\\ \href{snowmass21.org/rare/blv}{Rare Processes and Precision Measurements Frontier: Topical Group on Baryon \& Lepton Number Violation} (RP4)
\end{center}
\fi

\noindent

The US particle physics community is preparing to identify and rank scientific priorities with the goal of shaping the physics program for the next few decades to confront critically important questions such as the lack of antimatter in the universe and the particle nature of dark matter as part of the Snowmass process. We identify several key opportunities which can address under-explored processes that can provide explanations for these phenomena. Members of the theoretical and experimental communities were brought together to discuss the challenges, potential, and ramifications for the detection of baryon number violation in the coming decade as part of a workshop ``Theoretical Innovations for Future Experiments Regarding Baryon Number Violation'', the first fully-virtual workshop hosted by the Amherst Center for Fundamental Interactions (ACFI), held August 3-6, 2020. 
The workshop ``Prospects for Baryon Number Violation by Two Units'' was originally organized at the ACFI, intended to be held April 2--4, 2020, but was regrettably cancelled due to the coronavirus pandemic. The virtual workshop was organized as part of the Snowmass 2021 Rare Processes and Precision Measurements Frontier, serving the Topical Group on Baryon \& Lepton Number Violation. The goal of this important and timely workshop was to report on the state of the field, survey the opportunities in experiment and theory, and outline a path forward. Letters of Interest reflecting the community's input collected during the workshop were organized and submitted as part of the Snowmass process.
%important and timely workshop
%A summertime workshop frame ensures that we can better serve the experimental community interested in BNV by surveying the opportunities in experiment and theory in order to build the foundation for the strategy for Snowmass. 
%This will allow for in-depth planning for both topics and convener strategies in order to be noticed within the broader field.
%The workshop was open for all to attend, most joining from 10am-2pm EDT, August 3rd-6th, 2020 via Zoom. An optional Snowmass LOI/Contributed Paper writing session was also operated everyday of the workshop, typically from 2pm-3pm. 
%Abstract submission was limited to previously agreed participants in light of postponed plans and time limits, and though several prospective contributors were unable at the last minute, all talk slots were filled with their previously specified topics.

The overarching topic of this workshop is the violation of Baryon-minus-Lepton ($ \mathcal{B} - \mathcal{L} $) number. $ \mathcal{B} - \mathcal{L} $ number is exactly conserved in the Standard Model, but the observed matter-antimatter asymmetry of the universe hints that beyond the Standard Model $ \mathcal{B} - \mathcal{L} $ violating processes could exist. Proton decay (PDK) experiments set very strong limits on $ \mathcal{B} $-violating interactions (though most conserve $ \mathcal{B} - \mathcal{L} $), pointing towards very high-energy scales around $10^{13}\,$TeV. However, there are models where the proton is stable while $\mathcal{B}$ is still not a good symmetry; for instance, if $ \mathcal{B} $ is only violated by two units,  $|\Delta\mathcal{B}|=2$. Such models lead to unique and powerful experimental signatures such as the transformations of neutrons into antineutrons ($n\rightarrow\bar{n}$), which are similar to kaon-antikaon oscillations due to strangeness-changing weak interactions, or decays of otherwise stable nuclei via dinucleon annihilation.

Recent years have seen significant theoretical developments of various aspects of these intriguing scenarios, and models have been developed that naturally avoid PDK limits while solving other problems within the Standard Model such as the matter-antimatter asymmetry of the universe. Lattice-QCD calculations have made tremendous improvements in calculating QCD matrix elements that connect $\mathcal{B}$-violating quark interactions to observables. Studies in effective field theories for $\mathcal{B}$-violating nuclear interactions have been initiated and applied to light nuclei, while novel intranuclear simulations have been developed to assess whether dinucleon decay processes can be separated from background in medium-heavy nuclei. 

At the same time, the prospects for experiments are compelling. With an increased data set and enhanced signal to background discrimination, Super-Kamiokande is poised to deliver the world's best limit on $n\rightarrow\bar{n}$ oscillations. Future facilities such as DUNE %PNPI Gatchina, 
and Hyper-Kamiokande are all expected to attain significantly increased sensitivities to $ \mathcal{B} - \mathcal{L} $ violation from intranuclear searches. A unique opportunity for substantially increased neutron flux at the European Spallation Source would enable a free $n\rightarrow\bar{n}$ oscillation search with $10^3 \times$ better sensitivity than the previous experiment, taking advantage of several decades of technological developments. Research and development for this effort is possible at existing facilities such as at Oak Ridge National Laboratory, while accessing complementary and relatively unexplored physics via neutrons coupling to a dark sector.

%Despite these recent exciting developments, the collective particle, lattice-QCD, nuclear, and experimental communities are currently rather disjoint and do not meet very often (if at all) to discuss strategy and theoretical necessities for mutualistic progress in the field. 
This workshop collected representatives from across multiple communities, including particle and nuclear physics, theory, phenomenology, and experiment to identify the major challenges and explore the prospects for discovering $|\Delta\mathcal{B}|=2$ violation in future experiments, while discussing the potential interpretation(s) of future experimental signals, or lower limits, in the broader context of $ \mathcal{B} - \mathcal{L} $ violation. This workshop builds off of others previously held at the \href{https://www.int.washington.edu/PROGRAMS/17-69W/}{INT}, \href{https://indico.ill.fr/indico/event/87/}{ILL}, and \href{https://indico.fysik.su.se/event/6570/}{NORDITA} exceedingly well.

\begin{center}
    ***
\end{center}
\noindent 
We gratefully acknowledge the Amherst Center for Fundamental Interactions at the University of Massachusetts Amherst and the Laboratory Directed Research and Development Program of Oak Ridge National Laboratory, managed by UT-Battelle, LLC, for the U. S. Department of Energy for supporting this workshop, and the Snowmass 2021 Rare Processes and Precision Measurement Frontier for hosting the workshop Indico webpage.
\normalsize

\vspace{10pt}

\noindent
%Co-organizers Include:\\
Joshua Barrow (University of Tennessee Knoxville) \\
%, \href{mailto:jbarrow3@vols.utk.edu}{jbarrow3@vols.utk.edu})\\
Leah Broussard (Oak Ridge National Laboratory) \\
%, \href{mailto:broussardlj@ornl.gov}{broussardlj@ornl.gov})\\
Jordy de Vries (University of Massachusetts Amherst/Riken Brookhaven) \\
%, \href{mailto:jdevries@umass.edu}{jdevries@umass.edu})\\
Michael Wagman (Fermi National Accelerator Laboratory)\\
%, \href{mailto:mwagman@fnal.gov}{mwagman@fnal.gov})\\
\textit{--The Organizing Committee}

\mychapter{0}{Workshop Program}

\renewcommand{\tabcolsep}{0.35cm}
\definecolor{LightCyan}{rgb}{0.88,1,1}

\begin{center}\Large
\hyphenpenalty=10000
\begin{tabular}{| c |p{3 cm} p{3 cm} p{3 cm} p{3 cm}| }
 \hline
 & Aug 3 & Aug 4 & Aug 5 & Aug 6 \\ 
 \hline
 10:00 & Rabindra Mohapatra & Bingwei Long & James Wells & Susan Gardner \\
 10:30 & Robert Shrock & Jean-Marc Richard & Alexey Fomin & Julian Heeck \\
 \hline
\rowcolor{LightCyan}
 11:00 & \multicolumn{4}{c|}{Coffee Break}  \\
 \hline
 11:30 & Valentina Santoro & Discussion & Zurab Berezhiani & Albert Young \\
 \hline
 \rowcolor{LightCyan}
 12:00 & \multicolumn{4}{c|}{Lunch Break}  \\
 \hline
 12:30 & Linyan Wan & Michael Wagman & Yuri Kamyshkov & Sudhakantha Girmohanta \\
 13:00 & Joshua Barrow & Bhupal Dev & Marcel Demarteau & David McKeen \\ 
 13:30 & William Snow & K. S. Babu & Discussion & Closeout  \\
 \hline
 \rowcolor{LightCyan}
 14:00 & \multicolumn{4}{c|}{Writing Session } \\
 \hline
\end{tabular}
\end{center}

\noindent
\href{https://www.physics.umass.edu/acfi/seminars-and-workshops/theoretical-innovations-for-future-experiments-regarding-baryon-number}{Workshop Website}\\
\href{https://www.physics.umass.edu/acfi/seminars-and-workshops/prospects-for-baryon-number-violation-by-two-units}{Previous Workshop Website} (postponed due to COVID-19)\\
\href{https://indico.fnal.gov/event/44472/}{Workshop Indico Site}

\noindent
\\
Below we list the speakers and the titles of their contributions as they appeared in the Workshop.
\section*{Monday, August 3\textsuperscript{rd}, 2020}

\begin{center}
\begin{tabular}{p{2 cm} p{15 cm} }
 10:00 -- 10:30 & \textit{Overview of some recent theoretical developments in neutron oscillation} \\
 & Rabindra Mohapatra \\
 10:30 -- 11:00 & \textit{Some Recent Results on Models with $n-\bar n$ Oscillations} \\
 & Robert Shrock \\
 
 11:00 -- 11:30 & Coffee Break \\
 
 11:30 -- 12:00 & \textit{The European Spallation Source and Future Free Neutron Oscillations Searches} \\
 & Valentina Santoro\\
 \\
 12:00 -- 12:30 & Lunch \\
 \\
 12:30 -- 13:00 & \textit{Neutron-antineutron oscillation search at Super-Kamiokande} \\
 & Linyan Wan\\
 13:00 -- 13:30 & \textit{Search for $n\rightarrow\bar{n}$ in the Deep Underground Neutrino Experiment} \\
 & Joshua Barrow \\
 13:30 -- 14:00 & \textit{Possible Use Of Neutron Optics for Optimization of a Free Neutron-Antineutron Oscillation Search} \\
 & William Snow \\
 
 14:00 -- 16:00 & Writing Session
\end{tabular}
\end{center}

\newpage %prevent ugly section header separation

\section*{Tuesday, August 4\textsuperscript{th}, 2020}

\begin{center}
\begin{tabular}{p{2 cm} p{15 cm} }
 10:00 -- 10:30 & \textit{Baryon-number violation by two units in chiral effective field theory} \\
 & Bingwei Long \\
 10:30 -- 11:00 & \textit{Calculation of the Suppression Factor for Bound Neutron-Antineutron Transformation} \\
 & Jean-Marc Richard \\
 
 11:00 -- 11:30 & Coffee Break \\
 
 11:00 -- 11:30 & Discussion \\
 \\
 12:00 -- 12:30 & Lunch \\
 \\
 12:30 -- 13:00 & \textit{Lattice QCD matrix elements of $|\Delta \mathcal{B}| = 2$ operators} \\
 & Michael Wagman\\
 13:00 -- 13:30 & \textit{Update on the post-sphaleron baryogenesis model prediction for neutron-antineutron oscillation time} \\
 & Bhupal Dev \\
 13:30 -- 14:00 & \textit{Probing High Scale Theories with $n\rightarrow\bar{n}$ Oscillations} \\
 & K.\ S.\ Babu \\
 
 14:00 -- 16:00 & Writing Session
\end{tabular}
\end{center}

\section*{Wednesday, August 5\textsuperscript{th}, 2020}

\begin{center}
\begin{tabular}{p{2 cm} p{15 cm} }
 10:00 -- 10:30 & \textit{Neutron-antineutron oscillation improvements and baryogenesis} \\
 & James Wells \\
 10:30 -- 11:00 & \textit{Search for Neutron-Antineutron Oscillations with UCN} \\
 & Alexey Fomin \\
 
 11:00 -- 11:30 & Coffee Break \\
 
 11:30 -- 12:00 & \textit{New scenario for the neutron--antineutron oscillation: shortcut through mirror world} \\
 & Zurab Berezhiani \\
 \\
 12:00 -- 12:30 & Lunch \\
 \\
 12:30 -- 13:00 & \textit{Search for neutron oscillations to a sterile state ($ n \rightarrow n^{'} $) and to an antineutron ($ n \rightarrow \overline{n} $)} \\
 & Yuri Kamyshkov\\
 13:00 -- 13:30 & \textit{Neutrons at ORNL and ESS: A Synergistic Program} \\
 & Marcel Demareatu \\
 11:00 -- 11:30 & Discussion \\
 
 14:00 -- 16:00 & Writing Session
\end{tabular}
\end{center}

\section*{Thursday, August 6\textsuperscript{th}, 2020}

\begin{center}
\begin{tabular}{p{2 cm} p{15 cm} }
 10:00 -- 10:30 & \textit{Searches for scalars that carry $\mathcal{B}$ or $\mathcal{L}$, taken broadly: whither and wherefore} \\
 & Susan Gardner \\
 10:30 -- 11:00 & \textit{Covering baryon number violation with inclusive searches} \\
 & Julian Heeck \\
 
 11:00 -- 11:30 & Coffee Break \\
 
 11:30 -- 12:00 & \textit{Measurements of Neutron Coupling to a Mirror Sector Using Spin Precession} \\
 & Albert Young \\
 \\
 12:00 -- 12:30 & Lunch \\
 \\
 12:30 -- 13:00 & \textit{Exciting New Possibilities for Baryon Number Violation} \\
 & Sudhakantha Girmohanta\\
 13:00 -- 13:30 & \textit{Perspectives on Baryon Number Violation} \\
 & David McKeen \\
 13:30 -- 14:00 & Closeout \\
 
 14:00 -- 16:00 & Writing Session
\end{tabular}
\end{center}

\mychapter{0}{Registered Participant List}

\label{sec:Participants}
\begin{center}
\hyphenpenalty=10000
\begin{tabular}{p{3.25 cm} p{6.75 cm} p{7 cm} }
K.\ S.\ Babu & Oklahoma State University & \href{mailto:babu@okstate.edu}{babu@okstate.edu} \\
Joshua Barrow & The University of Tennessee & \href{mailto:jbarrow3@vols.utk.edu}{jbarrow3@vols.utk.edu} \\
Zurab Berezhiani & Universit{\'a} di L’Aquila \& Gran Sasso National Laboratory & \href{mailto:zurab.berezhiani@aquila.infn.it}{zurab.berezhiani@aquila.infn.it}  \\
Leah Broussard & Oak Ridge National Laboratory & \href{mailto:broussardlj@ornl.gov}{broussardlj@ornl.gov}  \\
Marcel Demarteau & Oak Ridge National Laboratory & \href{mailto:demarteau@ornl.gov}{demarteau@ornl.gov} \\
Bhupal Dev & Washington University in St. Louis & \href{mailto:bdev@physics.wustl.edu}{bdev@physics.wustl.edu} \\
Alexey Fomin &  National Research Center ``Kurchatov Institute" - Petersburg Nuclear Physics Institute  & \href{mailto:fomin\_ak@pnpi.nrcki.ru}{fomin\_ak@pnpi.nrcki.ru} \\
Susan Gardner & University of Kentucky & \href{mailto:svg@pa.uky.edu}{svg@pa.uky.edu}  \\
Sudhakantha Girmohanta & Stony Brook University & \href{mailto:sudhakantha.girmohanta@stonybrook.edu}{sudhakantha.girmohanta@stonybrook.edu} 	\\
Elena Golubeva & Institute for Nuclear Research, Russian Academy of Sciences, Moscow & \href{mailto:golubeva@inr.ru}{golubeva@inr.ru} \\
Maury Goodman & Argonne National Laboratory & \href{mailto:maury.goodman@anl.gov}{maury.goodman@anl.gov} \\
Julian Heeck & University of Virginia & \href{mailto:heeck@virginia.edu}{heeck@virginia.edu} \\
Yeon-jae Jwa & Columbia University & \href{mailto:yj2429@fnal.gov }{yj2429@fnal.gov}\\
Yuri Kamyshkov & University of Tennessee & \href{mailto:kamyshkov@utk.edu}{kamyshkov@utk.edu} \\
Georgia Karagiorgi & Columbia University & \href{mailto:georgia@nevis.columbia.edu}{georgia@nevis.columbia.edu}  \\
Praveen Kumar & The University of Sheffield & \href{mailto:pkumar3@sheffield.ac.uk }{pkumar3@sheffield.ac.uk} \\
Bingwei Long & Sichuan University & \href{mailto:bingwei@scu.edu.cn}{bingwei@scu.edu.cn} \\
Prajwal M. Murthy & University of Chicago & \href{mailto:	prajwal@mohanmurthy.com}{	prajwal@mohanmurthy.com} \\
Rukmani Mohanta & University of Hyderabad & \href{mailto:rukmani98@gmail.com}{rukmani98@gmail.com} \\
David McKeen & TRIUMF & \href{mailto:mckeen@triumf.ca}{mckeen@triumf.ca}  \\
Rabindra Mohapatra & University of Maryland & \href{mailto:rmohapat@umd.edu}{rmohapat@umd.edu} \\
Pavel Fileviez Perez & Case Western Reserve University &  \href{mailto:pxf112@case.edu}{pxf112@case.edu} \\
Jean-Marc Richard & IN2I, IN2P3 \& Universit\'e de Lyon  & 	\href{mailto:j-m.richard@ipnl.in2p3.fr}{j-m.richard@ipnl.in2p3.fr}  \\
Enrico Rinaldi & Arithmer Inc. \& RIKEN iTHEMS & \\
Valentina Santoro &	European Spallation Source & \href{mailto:valentina.santoro@ess.eu}{valentina.santoro@ess.eu} \\
Robert Shrock & C. N. Yang Institute for Theoretical Physics, Stony Brook University	& \href{mailto:robert.shrock@stonybrook.edu}{robert.shrock@stonybrook.edu}  \\
William Michael Snow & Indiana University & \href{mailto:wsnow@indiana.edu}{wsnow@indiana.edu}  \\
Anca Tureanu & University of Helsinki & \href{mailto:anca.tureanu@helsinki.fi}{anca.tureanu@helsinki.fi} \\
Jordy de Vries & University Massachusetts Amherst \& RIKEN BNL Research Center & 	\href{mailto:jdevries@umass.edu}{jdevries@umass.edu} \\
Michael Wagman & Fermi National Accelerator Laboratory &  \href{mailto:mwagman@fnal.gov }{mwagman@fnal.gov }\\
Linyan Wan & 	Boston University & \href{mailto:wanly@bu.edu }{wanly@bu.edu} \\
James Wells & University of Michigan &  \href{mailto:jwells@umich.edu }{jwells@umich.edu} \\
Sze Chun Yiu & Stockholms Universitet & \href{mailto:sze-chun.yiu@fysik.su.se}{sze-chun.yiu@fysik.su.se} \\
Albert Young & North Carolina State University \& Triangle Universities Nuclear Laboratory & \href{mailto:aryoung@ncsu.edu}{aryoung@ncsu.edu}
\end{tabular}
\end{center}

\tableofcontents
\markboth{}{}
\mainmatter
%\mychapter{0}{Executive Summary}

\mychapter{1}{Workshop Contributions}
\section{Theoretical Overviews}

\subsection[Overview of some recent theoretical developments in neutron oscillations\\ \small\textit {Rabindra Mohapatra}]{\href{https://indico.fnal.gov/event/44472/contributions/192036/}{Overview of some recent theoretical developments in neutron oscillations}}
\chapterauthor{Rabindra Mohapatra \\ University of Maryland \\ E-mail: \href{mailto:rmohapat@physics.umd.edu}{rmohapat@physics.umd.edu}}
%\subsection{Rabindra Mohapatra (\href{mailto:rmohapat@physics.umd.edu}{rmohapat@physics.umd.edu})\\
%\href{https://indico.fnal.gov/event/44472/contributions/192036/attachments/132300/162499/ACFI_2020_final.pdf}{Overview of some recent theoretical developments in neutron oscillations}}
There are a number of puzzles of beyond the standard model physics that can be probed directly by the process of neutron-anti-neutron oscillation in contrast with the other popular baryon violating process i.e. the typical GUT motivated proton decay mode $p\rightarrow e^+ \pi^0$. The most important of them is a direct understanding of the baryon asymmetry of the universe on which the typical GUT motivated baryon violation cannot. Also if neutron oscillation is observable, leptogenesis mechanism also does not work. The mechanism for such baryogenesis is the post sphaleron model which implemented in the context of $SU(2)_L \times SU(2)_R \times SU(4)_C$  model for neutron oscillation leads to an upper limit on neutron-antineutron oscillation time within the reach of currently proposed experiments. Furthermore, if neutrino-less double beta decay fails to yield a positive signal, an alternative way to establish that lepton number is violated and neutrinos are their own antiparticles is to discover both proton decay and neutron oscillations. Also the belief that neutrinos are likely to be Majorana fermions strongly suggests that there may be a small Majorana component to the neutron mass which leads to neutron oscillation. All these arguments provide strong arguments for a new search for neutron-antineutron oscillation. In the second part of the talk, I point out some constraints arising from big bang nucleosynthesis that suppress the neutron mirror neutron oscillation which is under study in several experiments.

\subsection[Some Recent Results on Models with $n\rightarrow\bar{n}$ Oscillations\\ \small\textit {Robert Shrock}]{\href{https://indico.fnal.gov/event/44472/contributions/192553/}{Some Recent Results on Models with $n\rightarrow\bar{n}$ Oscillations}}
\chapterauthor{Robert Shrock \\ Stony Brook University \\ E-mail: \href{mailto:robert.shrock@stonybrook.edu}{robert.shrock@stonybrook.edu}}
%\subsection{Robert Shrock (\href{mailto:robert.shrock@stonybrook.edu}{robert.shrock@stonybrook.edu})\\
%\href{https://indico.fnal.gov/event/44472/contributions/192553/attachments/132305/162508/nnba.pdf}{Some Recent Results on Models with $n\rightarrow\bar{n}$ Oscillations}}
The violation of baryon number ($\mathcal{B}$) is a feature of many theories of fundamental physics going beyond the Standard Model (SM) and is expected as a requirement for explaining the observed baryon number asymmetry in the universe.  Because proton decay and $\mathcal{B}$-violating (BNV) decays of neutrons bound in nuclei are mediated by four-fermion operators with coefficients of the form $1/({\rm mass})^2$ (in four spacetime dimensions), while $n\rightarrow\bar{n}$ oscillations are mediated by six-quark operators with coefficients of the form $1/({\rm mass})^5$, one might naively think that BNV nucleon decays would be a more important manifestation of baryon number violation than $n\rightarrow\bar{n}$ transitions. However, there are models in which BNV nucleon decays are either absent or can be suppressed far below experimental limits, so that $n\rightarrow\bar{n}$ transitions are the dominant manifestation of baryon number violation.  Here we discuss a class of models of this type, in which $n\rightarrow\bar{n}$ oscillations can occur at observable levels.  These are extra-dimensional theories with SM fermions having localized wave functions in the extra dimensions. The suppression of proton and bound neutron decays is achieved by separating the wave function centers of the quarks from those of the leptons by a sufficiently large distance in the extra dimensions.  However, this does not suppress $n\rightarrow\bar{n}$ transitions, since the operators mediating these transitions do not involve leptons. Thus, in these theories, $n\rightarrow\bar{n}$ oscillations and the associated $\Delta \mathcal{B}=-2$ dinucleon decays can be the dominant manifestation of baryon-number violation. Starting from the underlying theory in the higher-dimensional space, one integrates relevant operator products over the extra dimensions to obtain the low-energy effective Lagrangian in four spacetime dimensions.  Analyses are given within the context of the SM gauge group and a left-right-symmetric gauge group. An interesting feature of the left-right symmetric model is that certain six-quark operators are not suppressed by exponential factors arising from this integration over the extra dimensions.  This means that, for a given mass scale $M_{n \bar n}$ characterizing the BNV physics responsible for the $n\rightarrow\bar{n}$ transitions, these could occur at a larger rate in the left-right symmetric model than in the model with a SM gauge symmetry.  These results provide further motivations for new experiments to search for $n\rightarrow\bar{n}$ transitions and associated dinucleon decays.  Our related publications include those by Nussinov and Shrock\cite{Nussinov:2001rb,Nussinov:2020wri}, along with Girmohanta and Shrock \cite{Girmohanta:2019fsx,Girmohanta:2020qfd,Girmohanta:2019cjm}.

\newpage 

\subsection[Baryon-number violation by two units in chiral effective field theory \\ \small\textit {Bingwei Long}]{\href{https://indico.fnal.gov/event/44472/contributions/192552/}{Baryon-number violation by two units in chiral effective field theory}}
\chapterauthor{Bingwei Long \\ Sichuan University\\ E-mail: \href{mailto:bingwei@scu.edu.cn}{bingwei@scu.edu.cn} \\ Femke Oosterhof and Rob Timmermans \\ University of Groningen  \\ Jordy de Vries \\ University of Massachusetts, Amherst, and RIKEN BNL Research Center \\ \textit{and} \\ Ubirajara van Kolck \\ Laboratoire Ir\`ene Joliot-Curie and University of Arizona
}
%\subsection{Bingwei Long (\href{mailto:bingwei@scu.edu.cn}{bingwei@scu.edu.cn})\\
%\href{https://indico.fnal.gov/event/44472/contributions/192552/attachments/132496/162882/BingweiLong_nnbar_acfi.pdf}{Baryon-number violation by two units in chiral effective field theory}}

When studying physics beyond the Standard Model (BSM) at the intensity frontier, one often looks for possible tiny BSM signals in the backdrop of atomic nuclei. Baryon number violating physics is one of the examples, and more specifically, my contribution focuses on baryon number violation by two units, $|\Delta \mathcal{B}| = 2$. The model-independent framework I will discuss is a tower of effective field theories that begins with higher dimensional operators to extend the Standard Model and ends with with chiral effective field theory equipped with $|\Delta \mathcal{B}| = 2$ hadronic operators. 

The deuteron lifetime is used as an application to illustrate this framework. The smallness of the deuteron binding momentum makes it possible to treat pion exchanges as perturbations. This in turn allows for an analytic expression up to next-to-leading order that links $n\rightarrow\bar{n}$ oscillation time to the deuteron lifetime, with several nucleon-nucleon and nucleon-antinucleon scattering parameters describing the Standard Model physics involved. It highlights quantitatively why nuclei with a loosely bound neutron could be sensitive to neutron-antineutron oscillation time. The other emphasis is given to how a consistent power counting is built and what statement on theoretical uncertainty can be drawn out of it.

\subsection[Calculation of the Suppression Factor for Bound Neutron-Antineutron Transformation\\ \small\textit {Jean-Marc Richard}]{\href{https://indico.fnal.gov/event/44472/contributions/192551/}{Calculation of the Suppression Factor for Bound Neutron-Antineutron Transformation}}
\chapterauthor{Jean-Marc Richard \\ Universit\'e de Lyon \& IN2P3 \\ E-mail: \href{mailto:j-m.richard@ipnl.in2p3.fr}{j-m.richard@ipnl.in2p3.fr}}
%\subsection{Jean-Marc Richard (\href{mailto:jmrichar@ipnl.in2p3.fr}{jmrichar@ipnl.in2p3.fr})\\
%\href{https://indico.fnal.gov/event/44472/contributions/192551/attachments/132289/162477/Snowmass-2020-Richard.pdf}{Calculation of the Suppression Factor for Bound Neutron-Antineutron Transformation}}
%I review and revisit the calculation of the lifetime of nuclei due to neutron-antineutron oscillations.
%It is  stressed that the oscillation and the subsequent annihilation take place mainly outside the nucleus and thus hardly suffer from drastic renormalization due to the nuclear medium. The ingredients of the calculation can be safely extracted from nuclear shell-model wave-functions, and optical models fitting the low-energy data on antinucleon-nucleus interaction. The main result is that the lifetime of a nucleus behaves as $T=T_R\,\tau_{n\bar n}^2$, with a factor $T_R$, often referred to as \textit{reduced lifetime} or \textit{suppression factor} of about $10^{22-23}\,$s$^{-1}$. A remarkable feature is that $T_R$ is stable against variations of the antinucleon-nucleus potential.
%%%%%%%%%%%%%% new file Oct 2, 2020
The lifetime of deuterium due to neutron-antineutron oscillations has been estimated by Sandars in 1980 \cite{Sandars:1980pr}, and recently revisited in the framework of chiral effective theories \cite{Oosterhof:2019dlo,Haidenbauer:2019fyd}. 
The formalism has been generalized by Dover et al.~\cite{Dover:1982wv} to heavier nuclei, using an effective shell model, while other appraoches were developed~\cite{Alberico:1982nu}. 

For each neutron shell of a $^{A}_Z X$ nucleus corresponding to an orbital momentum $\ell$ and a reduced radial function $u(r)$, there is an associated antineutron component $w(r)$ given by the Sternheimer equation
$$
 \frac{\hbar^2}{2\,m}\left[-w''(r)+\frac{\ell(\ell+1)}{r^2}\,w(r)\right]+ W(r)\,w(r)-E_{n,\ell}\,w(r)= \gamma\,u(r)~.
$$
where $m$ is the reduced mass of the neutron and the rest of the nucleus, $E_{n,\ell}$ the effective energy of the neutron on its shell, $\gamma=\hbar/\tau_{n\bar n}$ is the strength of the elementary $n\rightarrow \bar n$ transition, and  $W(r)$ the complex optical potential describing the interaction of an antineutron and the $^{A-1}_{\phantom{A-}Z} X$ nucleus, which is determined by a fit to the data on antinucleon-nucleus scattering experiments and antiprotonic atoms. Once this equation is solved, one gets the contribution of each shell to the annihilation width of the nucleus, as well as the spatial distribution of the antineutron. 

The first  result is that the lifetime of a nucleus behaves as $T=T_R\,\tau_{n\bar n}^2$, with a factor $T_R$, often referred to as \textit{reduced lifetime} or \textit{suppression factor} of about $10^{22-23}\,$s$^{-1}$. Another  feature is that $T_R$ is remarkably stable against variations of the antinucleon-nucleus potential. One should also stress that the oscillation and the subsequent annihilation take place mainly outside the nucleus and thus hardly suffer from drastic renormalization due to the nuclear medium.

\begin{figure}[h!]
 \centering
 \includegraphics[width=.5\textwidth]{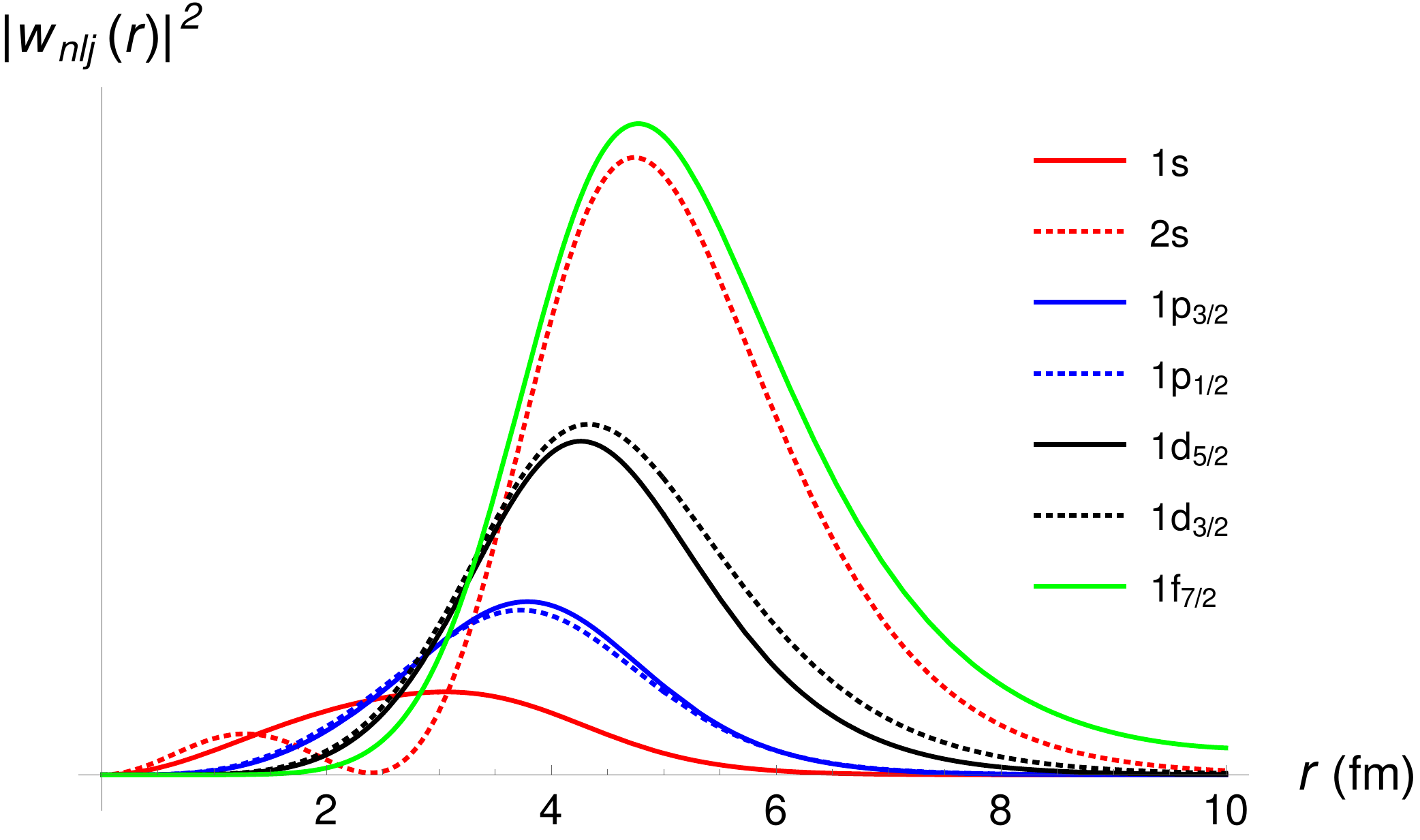}
 % Argon-nbar-dens.pdf: 600x355 px, 72dpi, 21.17x12.52 cm, bb=0 0 600 355
 \vspace*{.2cm}
 \caption{Antineutron density generated by the various neutron shells of ${}^{40}_{18}$Ar}
 \label{fig:richard-fig1}
\end{figure}
The first calculations were devoted to nuclei relevant for the early underground experiments. It has been recently updated for ${}^{40}_{18}$Ar of the DUNE experiment~\cite{Barrow:2019viz}. In Fig.~\ref{fig:richard-fig1} is shown a comparison of the antineutron densities generated by the various shells. Clearly the external neutrons contribute more than the internal ones, and the antineutrons are mostly outside the nucleus, so that their annihilation takes place at the surface.

\subsection[Lattice QCD matrix elements of $\Delta \mathcal{B} = 2$ operators\\ \small\textit {Michael Wagman and Enrico Rinaldi}]{\href{https://indico.fnal.gov/event/44472/contributions/193469/}{Lattice QCD matrix elements of $\Delta \mathcal{B} = 2$ operators}}
\chapterauthor{Michael Wagman \\ Fermi National Accelerator Laboratory \\ E-mail: \href{mailto:mwagman@fnal.gov}{mwagman@fnal.gov}  \\ Enrico Rinaldi \\ Arithmer Inc. \& RIKEN iTHEMS \\ Sergey Syritsyn \\ Stony Brook University \& RIKEN BNL Research Center \\ \textit{and}\\ Michael Buchoff, Chris Schroeder, and Joseph Wasem \\ Lawrence Livermore National Laboratory  }
%\subsection{Michael Wagman (\href{mailto:mwagman@fnal.gov}{mwagman@fnal.gov})\\
%\href{https://indico.fnal.gov/event/44472/contributions/193469/attachments/132400/162666/Wagman_slides.pdf}{Lattice QCD matrix elements of $\Delta \mathcal{B} = 2$ operators}}
Theories of $\mathcal{(B-L)}$ violation beyond the Standard Model (BSM) generically lead to the appearance of six-quark operators in Standard Model effective field theory that give rise to neutron-antineutron oscillations and $|\Delta \mathcal{B}| = 2$ nuclear decays. It is possible to reliably connect the results of experimental searches for these processes to constraints on the parameters of BSM physics theories. However, this effort requires Standard Model calculations of the matrix elements of these six-quark operators between hadronic states, where non-perturbative physics gives important contributions.
The framework of lattice quantum chromodynamics (LQCD)
is a well-known tool to calculate hadronic physics quantities with controlled and improvable theoretical errors.
In this talk, I report LQCD calculations of the matrix elements of a complete basis of $|\Delta \mathcal{B}| = 2$ six-quark operators.
Moreover, I will show how these non-perturbative results compare to previous model calculations and what kind of implications we expect for current and future searches for $|\Delta \mathcal{B}| = 2$ processes (including nuclear decays).

\subsection[Update on the post-sphaleron baryogenesis model prediction for neutron-antineutron oscillation time \\ \small\textit {P. S. Bhupal Dev}]{\href{https://indico.fnal.gov/event/44472/contributions/192053/}{Update on the post-sphaleron baryogenesis model prediction for neutron-antineutron oscillation time}}
\chapterauthor{P. S. Bhupal Dev \\ Washington University in St. Louis \\ E-mail: \href{mailto:bdev@physics.wustl.edu}{bdev@physics.wustl.edu}}
%\subsection{P. S. Bhupal Dev (\href{mailto:bdev@physics.wustl.edu}{bdev@physics.wustl.edu})\\
%\href{https://indico.fnal.gov/event/44472/contributions/192053/attachments/132385/162674/dev_acfi20.pdf}{Update on the post-sphaleron baryogenesis model prediction for neutron-antineutron oscillation time}}
Post-sphaleron baryogenesis (PSB) is an attractive low-scale mechanism to explain the observed matter-antimatter asymmetry of the Universe. As the name suggests, the generation of baryon asymmetry occurs after the sphalerons have gone out of equilibrium. The same  $|\Delta \mathcal{B}|=2$ operator that gives rise to baryogenesis in this scenario also leads to $n\rightarrow\bar{n}$ oscillation, thus making an intimate connection between the two $\mathcal{B}$-violating observables. The PSB mechanism, when embedded in a quark-lepton unified model based on the Pati-Salam gauge group, leads to an absolute upper limit on the neutron-antineutron oscillation time, which might be within reach of future experiments. The upper bound on the $n\rightarrow\bar{n}$ oscillation time ($\tau_{n\bar{n}}$) is a consequence of the bounded range of the parameters that $\tau_{n\bar{n}}$ depends on. Requiring the model to reproduce the observed neutrino masses and oscillation parameters fixes the form of the coupling matrices involved in the generation of baryon asymmetry in this model. The given form of the matrix essentially puts an upper bound on the maximum baryon asymmetry $(\epsilon_{B}^{\rm max})$ that can be produced. This asymmetry is produced by the decay of a color-sextet real scalar $S$ to  quarks and anti-quarks, mediated by scalar diquarks $\Delta_{dd}$ and $\Delta_{ud}$. This process happens at a specific decay temperature ($T_d$) dependent on the mass of the scalar, $M_S$, and of the diquarks, $M_{\Delta_{ud}}$ and $M_{\Delta_{dd}}$. The observed baryon asymmetry $\eta_B$ at the recombination epoch is related to the baryon asymmetry produced in the decay of $S$ through a dilution factor proportional to $T_d/M_S$. Since the dilution cannot be more than $\eta_B/\epsilon_{B}^{\rm max}$, this sets an upper bound on the range of $M_S$. On the other hand, the $S$ must decouple while being relativistic before its decay produces the asymmetry, which requires the $\Delta_{ud}$ and $\Delta_{dd}$ masses to be at least a factor of 5-10 larger than $M_S$. If $\Delta_{ud}$ and $\Delta_{dd}$ are too heavy, however, this will drive the $T_d$ lower, and hence, a larger dilution. This delicate interplay between the model parameters makes this scenario quite predictive and testable. 
We present an updated prediction of the $n\rightarrow\bar{n}$ oscillation time, based on an  improved calculation of the baryon asymmetry, which takes into account both wave-function and vertex correction diagrams. We also make use of the recent lattice QCD results on the relevant $\Delta\mathcal{B}$ operators.  
Apart from $n\rightarrow\bar{n}$, this model also features multi-TeV-scale scalar diquarks, which could  be searched for at the LHC and future hadron colliders.

\subsection[Probing High Scale Theories with $n\rightarrow\bar{n}$ Oscillations \\ \small\textit {K. S. Babu}]{\href{https://indico.fnal.gov/event/44472/contributions/192068/}{Probing High Scale Theories with $n\rightarrow\bar{n}$ Oscillations} }
\chapterauthor{K. S. Babu \\ Oklahoma State University \\ E-mail: \href{mailto:babu@okstate.edu}{babu@okstate.edu}}
%\subsection{K. S. Babu (\href{mailto:babu@okstate.edu}{babu@okstate.edu})\\
%\href{https://indico.fnal.gov/event/44472/contributions/192068/attachments/132386/162647/Bequal2_Babu.pdf}{Probing High Scale Theories with $n\rightarrow\bar{n}$ Oscillations}}
Neutron-antineutron oscillations ($n\rightarrow\bar{n}$) can be used to probe theories at a high energy scale, such as grand unified theories. In this talk I will illustrate this with two examples.  In the first example, $n\rightarrow\bar{n}$ oscillation arises in a left-right symmetric model realized near the GUT scale that provides a solution to the strong CP problem.  The  $n\rightarrow\bar{n}$ oscillation time is closely tied to neutrino masses, and is expected to be in the range of $\tau_{n\bar{n}} \sim10^8-10^{10}$ sec. In the second example, $SO(10)$ grand unified theory breaks to the standard model directly, but leaves behind a color sextet scalar field at the TeV scale.  This scalar helps with unification of gauge couplings and leads to $n\rightarrow\bar{n}$ oscillations, which is closely tied to baryon asymmetry generation.  For typical values of the model parameters, $\tau_{n\bar{n}} \sim 10^9 - 10^{10}$ sec. is obtained.

\subsection[Neutron-antineutron oscillation improvements and baryogenesis\\ \small\textit {James Wells}]{\href{https://indico.fnal.gov/event/44472/contributions/193327/}{Neutron-antineutron oscillation improvements and baryogenesis}}
\chapterauthor{James Wells \\ University of Michigan \\ E-mail: \href{mailto:jwells@umich.edu}{jwells@umich.edu}}
%\subsection{James Wells (\href{mailto:jwells@umich.edu}{jwells@umich.edu})\\
%\href{https://indico.fnal.gov/event/44472/contributions/193327/attachments/132461/162842/ACFI_Wells19590.pdf}{Neutron-antineutron oscillation improvements and baryogenesis}}
Wherein I discuss how improvements on neutron-antineutron oscillations and its impact on a minimal theory of baryogenesis.

\subsection[New scenario for neutron--antineutron oscillations: a shortcut through a mirror sector \\ \small\textit {Zurab Berezhiani}] {\href{https://indico.fnal.gov/event/44472/contributions/193343/}{New scenario for neutron--antineutron oscillations: a shortcut through a mirror sector} }
\chapterauthor{Zurab Berezhiani \\ Universit{\'a} di L’Aquila \\ E-mail: \href{mailto:zurab.berezhiani@aquila.infn.it}{zurab.berezhiani@aquila.infn.it}}
%\subsection{Zurab Berezhiani (\href{mailto:zurab.berezhiani@aquila.infn.it}{zurab.berezhiani@aquila.infn.it})\\
%\href{https://indico.fnal.gov/event/44472/contributions/193343/attachments/132455/162828/Berezhiani-2020-B2-USA.pdf}{New scenario for the neutron--antineutron oscillation: shortcut  through mirror world}}

Existing bounds on the neutron-antineutron mass mixing, $\epsilon_{n\bar n} < {\rm few} \times 10^{-24}$~eV, impose a severe upper limit  on $n \rightarrow \bar n$ transition probability, $P_{n\bar n}(t) < (t/0.1 ~{\rm s})^2 \times 10^{-18}$ or so,  where $t$ is the neutron flight time. Here we propose a new mechanism of $n\rightarrow\bar{n}$ transition which is not induced by direct mass mixing $\epsilon_{n\bar n}$ but is mediated instead by the neutron  mass mixings $\epsilon_{nn'}$ and $\epsilon_{n\bar{n}'}$ with the hypothetical states of mirror neutron $n'$ and mirror antineutron $\bar{n}'$ which can be  as large as $\sim 10^{-14}$~eV or so, without contradicting present experimental limits and nuclear stability bounds. The probabilities of $n\rightarrow n'$ and $n\rightarrow\bar{n}'$ transitions, $P_{nn'}$ and $P_{n\bar{n}'}$, depend on environmental conditions in mirror sector, and by scanning over the magnetic field values in experiments they can be resonantly amplified. This opens up a possibility of $n\rightarrow\bar{n}$ transition with the probability $P_{n\bar n} = P_{nn'} \cdot P_{n\bar{n}'}$ which can reach the values up to $\sim 10^{-8} $. For finding this effect  in real experiments, the magnetic field  should not be suppressed but properly varied with small steps. This scenario points towards the scale of few TeV of new physics which can be responsible for these mixings, and can also suggest a new low scale co-baryogenesis mechanism between ordinary and mirror sectors.

\subsection[Searches for scalars that carry $\mathcal{B}$ or $\mathcal{L}$, taken broadly: whither and wherefore \\ \small\textit {Susan Gardner}]{\href{https://indico.fnal.gov/event/44472/contributions/193309/}{Searches for scalars that carry $\mathcal{B}$ or $\mathcal{L}$, taken broadly: whither and wherefore\footnote{I would like to thank Xinshuai Yan (xinshuai@mail.ccnu.edu.cn) for collaboration on the topics discussed here.}}}
\chapterauthor{Susan Gardner \\ University of Kentucky \\ E-mail: \href{mailto:gardner@pa.uky.edu}{gardner@pa.uky.edu}}
%\subsection{Susan Gardner (\href{mailto:gardner@pa.uky.edu}{gardner@pa.uky.edu})\\
%$$\href{https://indico.fnal.gov/event/44472/contributions/193309/attachments/132482/162865/svg_acfi_bnv_6aug20.pdf}{Searches for scalars that carry $\mathcal{B}$ or $\mathcal{L}$, taken broadly: whither and wherefore\footnote{I would like to thank Xinshuai Yan (xinshuai@mail.ccnu.edu.cn) for collaboration on the topics discussed here.}}}
%In many models of new physics, the expected rate of processes that break baryon number by two units rests on the features of a poorly known scalar sector, whose members can carry $\mathcal{B}$ or $\mathcal{L}$ quantum numbers. Thus the BNV discovery prospects in these channels are controlled by the extent to which the associated scalars are excluded by experiments. Working in the context of minimal scalar models, I will survey and discuss the existing constraints and note what windows of opportunity remain for the discovery of light new scalars. With these, new experiments become tenable, and I emphasize the complementary of these to other ongoing efforts and their broader implications.
The severity of the experimental limits on the non-observation of proton decay
and, generally, on 
the mediation of unobserved $|\Delta \mathcal{B}|=1$ and $|\Delta\mathcal{(B-L)}|=0$ processes via
$d=6$ operators, has long made it seem that 
%the scale of BNV is no less than 
BNV is physics of the GUT scale, with $\Lambda \simeq 10^{16}\, {\rm GeV}$, suggesting
that $|\Delta \mathcal{B}|=2$ processes, such as $n\rightarrow\bar{n}$ oscillations, mediated by $d=9$ operators, are very suppressed indeed. 
Yet we know this need not be the case, because predictive models have been constructed that ``look down''
from the GUT scale 
to explain not only the non-observation of proton
decay but also to predict 
%the possibility of 
observable $n\rightarrow\bar{n}$ oscillations~\cite{Babu:2012vc} --- and indeed there are other
TeV-scale models that act to a similar end~\cite{Babu:2006xc,Babu:2013yca,Dev:2015uca,Allahverdi:2017edd}.
%--- and there are more models that
%predict 
Nevertheless,
the idea that BNV (and also LNV) is intrinsically {\it very} high-energy physics has influenced experimental searches,
and it is possible that new experimental avenues for the detection of BNV and LNV by 2 units, through $d\ge 9$ operators, have been overlooked. 
In this vein it is useful to recall the
minimal scalar models that have been developed for the study of BNV (and LNV)
without proton decay~\cite{Arnold:2012sd,Gardner:2018azu},
%~\cite{Davies:1990sc,Bowes:1996xy,Arnold:2012sd,Arnold:2013cva,Gardner:2018azu}, 
for which only those
scalars that respect SM gauge symmetries, have interactions of mass dimension 4 or less,
and do not mediate $|\Delta \mathcal{B}|=1$ processes at tree level are considered. The scalar interactions
do not select a particular mass scale; rather, the allowed masses and couplings should be determined from
experiment, much as in searches for light hidden sectors~\cite{Alexander:2016aln}.
Certain combinations of these scalars
can give rise to $n\rightarrow\bar{n}$ oscillations, $\pi^-\pi^- \to e^- e^-$ decay
--- the leading chiral contribution to neutrinoless
double beta decay~\cite{Prezeau:2003xn} --- and to $e^- p \to e^+ {\bar p}$, e.g., 
arising from an effective $d=12$ operator; the observation
of any two of these processes would imply that the third also exists~\cite{Gardner:2018azu}. 
For example, the observation of
$e^- p \to e^+ {\bar p}$ and $n\rightarrow\bar{n}$
oscillations
would imply that $\pi^-\pi^- \to e^- e^-$ decay exists and imply that the
neutrino is Majorana~\cite{Gardner:2018azu}.
%In many models of new physics, the expected rate of processes that break baryon number by two units rests on the features of a poorly known scalar sector,\
% whose members can carry $\mathcal{B}$ or $\mathcal{L}$ quantum numbers.                                                                                    
%The BNV discovery prospects in these cases                                                                                                                 
Since these processes are all associated with effective operators of $d=9$ and higher,
their discovery prospects
%associated these higher dimension operators in these channels                                                                                              
are largely controlled by the extent to which the masses and couplings of the associated scalars are excluded by experiments.
The scalars of interest carry either $\mathcal{B}$ or $\mathcal{L}$, can be either
electroweak singlets or triplets, and can be color singlets, triplets, or sextets.
Flavor physics constraints severely
constrain colored scalars with intergenerational couplings, seemingly pushing
new scalars to high mass
scales~\cite{Arnold:2012sd}; but these can be evaded by choosing the flavor structure of the couplings. 
We suppose, say, that the scalar couplings act on first generation couplings only.
Even with such a choice, we note that experimental searches for light scalars, 
particularly those that carry electric charge and couple to leptons have been
extensive, and in what follows I summarize the 
%determination 
%we 
%review the extensive literature on that topic to determine 
%it is of particular interest to determine what 
%of 
the regions of 
%the pertinent scalar 
parameter space that have not yet been excluded~\cite{Dev:2018sel,Gardner:2019mcl}, noting 
that constraints on light scalars also bear on the possibility of
resolving the anomaly in the determination of the anomalous magnetic moment
of electron with light, lepton-number-carrying scalars~\cite{Gardner:2019mcl} --- 
and refer to that study for all details. 
%, and
%I recap the outcomes of that work. 
%and we review the results of that study. 
%In what follows we review 
%the study we made to determine whether sub-GeV scalar with lepton number could
%be used to solve 
%Thus we consider such scalars 
%as a particularly well-studied example. 
The doubly-charged scalars of interest are constrained just as
the doubly charged Higgs bosons
${H}_{L, R}^{\pm\pm}$~\cite{Mohapatra:1980yp} are. 
Indeed, there is an extensive collection of LEP data pertinent to
constraints on doubly charged scalars, and 
invariably the lower mass limits on
doubly charged scalars rely on the results from earlier experiments to exclude lighter mass candidates.
Ultimately limits from LEP on doubly charged scalars of less than about 25 GeV in mass come from precision
measurements of the $Z_0$ line shape. 
%Here we were able to 
For the scalar parameter space excluded in that way, we can extend the minimal scalar model to partially cancel that
contribution, enabling the survival of sub-GeV scalars to 
resolve the $(g-2)_e$ puzzle~\cite{Gardner:2019mcl}. 
Seemingly more stringent are studies
from CELLO at PETRA~\cite{LeDiberder:1988eu,Swartz:1989qz}, but in the $e^+e^- \to e^+e^- e^+e^-$ data they collected they required sharp vertices
to control backgrounds --- and these cuts remove 
any sensitivity that experiment could have had to
light, weakly coupled scalars. 
%solutions. 
Armed with this perspective, 
experiments to search for the processes mediated by $d\ge 9$ operators 
become tenable, and we look forward to new paths for the study of baryon-number-violating 
phenomena. 
%and I emphasize the complementary of these to other ongoing efforts and their broader implications.

\subsection[Covering baryon number violation with inclusive searches \\ \small\textit {Julian Heeck}]{\href{https://indico.fnal.gov/event/44472/contributions/192555/}{Covering baryon number violation with inclusive searches}}
\chapterauthor{Julian Heeck \\ University of Virginia \\ E-mail: \href{mailto:heeck@virginia.edu}{heeck@virginia.edu}}
%\subsection{Julian Heeck (\href{mailto:heeck@virginia.edu}{heeck@virginia.edu})\\
%\href{https://indico.fnal.gov/event/44472/contributions/192555/attachments/132481/162864/heeck-ACFI-2020.pdf}{Covering baryon number violation with inclusive searches}}
Baryon number violation is an extremely sensitive probe of physics beyond the Standard Model. However, the continued absence of any signals raises the question if we are actually looking in the right places or if we should broaden our search strategies. 
In the Standard Model effective field theory we find the first $\Delta \mathcal{L}$ operator at mass dimension $d=5$ (with $\Delta \mathcal{L}=2$) and the first $\Delta \mathcal{B}$ operators at $d=6$ (with $\Delta \mathcal{B} =\Delta \mathcal{L} =1$)~\citep{Weinberg:1980bf}. These should then be the dominating $\Delta \mathcal{L}$ and $\Delta \mathcal{B}$ operators, inducing processes such as neutrinoless double beta decays and two-body nucleon decays, respectively.

\begin{figure}[h]
\centering
\includegraphics[width=0.7\textwidth]{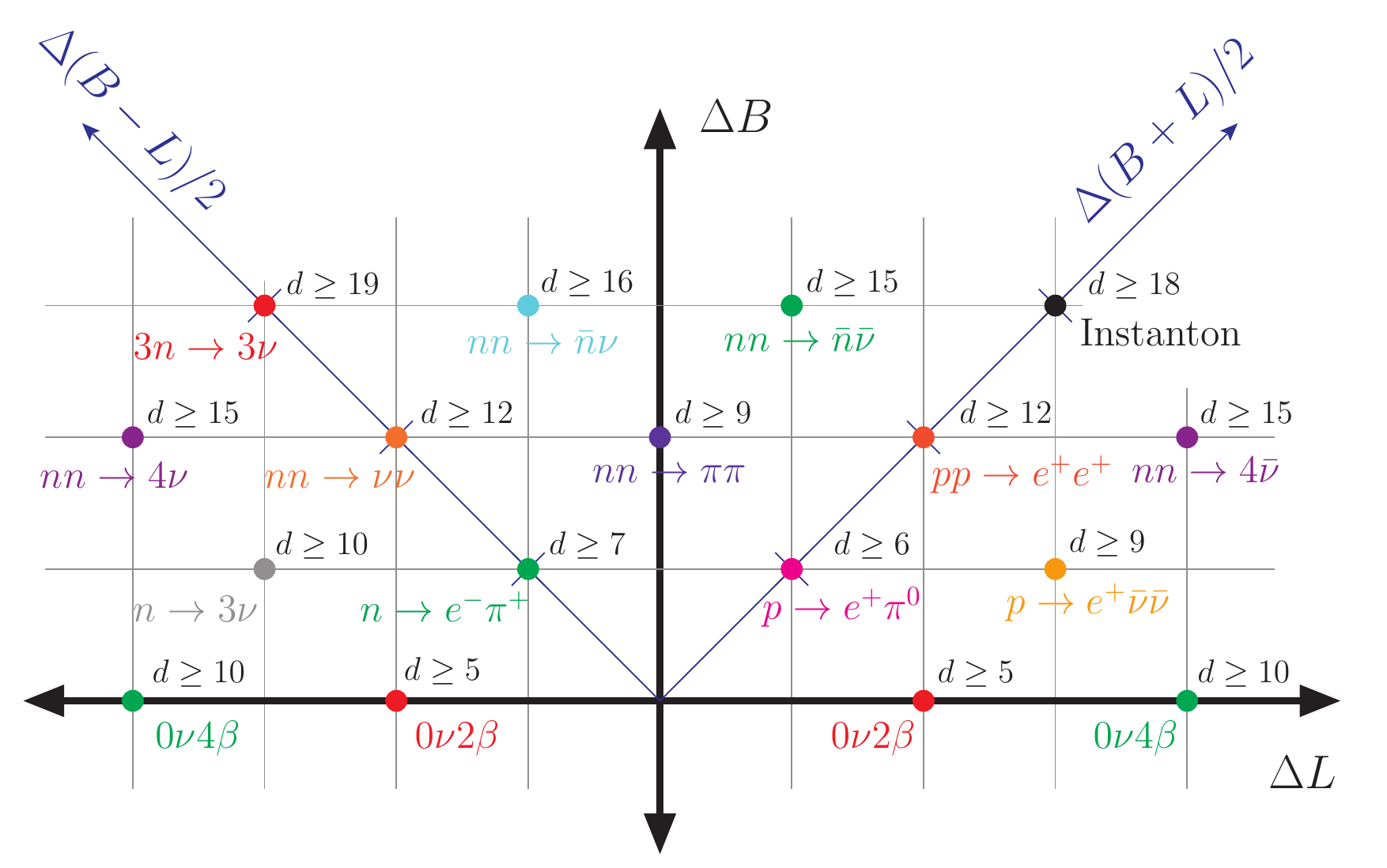}\vspace{0.6cm}
\caption{
Processes with baryon and lepton number violation by $\Delta \mathcal{B}$ and $\Delta \mathcal{L}$ units, respectively. We only show one example process, others are implied (e.g.~$nn\to\pi\pi$ also give $n$--$\bar n$ oscillation, $pp\to \pi^+\pi^+$, and many more). Also shown is the minimal mass dimension $d$ of the underlying effective operator. Adapted from Ref.~\cite{Heeck:2019kgr}.\vspace{0.6cm}
}
\label{fig:Heeck-fig1}
\end{figure}

This picture changes dramatically once we endow the effective field theory (or underlying models) with global or local symmetries contained in the Standard Model's global symmetry group $U(1)_{\mathcal{B}}\times U(1)_{\mathcal{L}}\times U(1)_{\mathcal{L}_\mu-\mathcal{L}_\tau}\times U(1)_{\mathcal{L}_\mu+\mathcal{L}_\tau-2 \mathcal{L}_e}$. It is not difficult to find or construct models in which these symmetries are only broken in one particular direction by a particular number of units, which then generates $\Delta \mathcal{L} $ and $\Delta \mathcal{B}$ processes that are entirely different from the standard search channels~\citep{Hambye:2017qix,Heeck:2019kgr}. Restricting ourselves to the $\Delta \mathcal{L}$ and $\Delta \mathcal{B}$ space, this is illustrated in Fig.~\ref{fig:Heeck-fig1}, but it should be kept in mind that lepton \emph{flavor} adds two more dimensions to this discussion~\citep{Hambye:2017qix}. For example, lepton flavor can single out nucleon decays such as $p\to e^-\mu^+\mu^+$~\citep{Hambye:2017qix} or $n\to K^+ \mu^+ e^-e^-$~\citep{Heeck:2019kgr} that carry $|\mathcal{L}_\alpha|>1$; the former has been constrained recently by Super-Kamiokande~\citep{Tanaka:2020emn}.

Nucleon decays are such a sensitive experimental probe that even $\Delta \mathcal{B}$ operators with mass dimension $d\gg 6$ could give observable signals. This makes it necessary to search for nuclear decays beyond the typical two-body proton decay modes in order to cover as much parameter and model space as possible. Dedicated \emph{exclusive} searches for multi-body nucleon decays or multi-nucleon decays would have the best sensitivity but are not a realistic option to cover all of the thousands of channels that are kinematically possible. A far more practical approach are \emph{inclusive} searches~\cite{Heeck:2019kgr}, e.g.~$p\to e^++\text{anything}$, which will not be as sensitive but are applicable to a wide variety of models, including light new particles~\cite{Heeck:2020nbq} and some dark-matter induced $\Delta \mathcal{B}$ signatures.
Existing limits from the inclusive searches $N\to e^++\text{anything}$ and $N\to \mu^++\text{anything}$ are of the order of $10^{30}\,$yr and $10^{31}\,$yr, respectively~\citep{Zyla:2020zbs}.
The vast amount of data already collected by Super-Kamiokande should make it possible to improve upon these 40-year-old limits~\cite{Heeck:2019kgr}.

One special case of inclusive searches comes in the form of invisible (multi-)neutron decay searches, e.g.~$n\to\text{neutrinos}$. In this case the final state is invisible and the only signature comes from the characteristic de-excitation radiation emitted by the daughter nucleus. Current limits are of the order of $10^{30}\,$yr~\cite{Araki:2005jt,Anderson:2018byx} but can probably be improved in future detectors such as JUNO, DUNE, and Hyper-Kamiokande~\cite{Heeck:2019kgr}.
These invisible neutron searches are important for full coverage and already provide most limits in Fig.~\ref{fig:Heeck-fig1} far away from the origin. One can easily prove that \emph{any} effective $\Delta \mathcal{B}$ operator generates decays of $|\Delta \mathcal{B}|$ neutrons into neutrinos, and even though these might not always be the dominant decay mode this makes it possible to use invisible searches to constrain \emph{every} model that leads to baryon number violation.
Furthermore, these invisible searches are also sensitive to new \emph{light} particles that are not covered by typical effective-field-theory discussions~\cite{Heeck:2020nbq}. For example, a new light scalar $\phi$ carrying $\mathcal{B=L}=1$ can lead to the decay $n\to \phi \bar{\nu}$, which is once again invisible~\cite{Heeck:2020nbq}.

The search for baryon number violation is well motivated and an extremely sensitive probe of physics at high energies, operators of high mass dimension, and processes with high multiplicity. To maximize the discovery potential of future detectors such as JUNO, DUNE, and Hyper-Kamiokande it is important to cast a wide net and search for all possible $\Delta \mathcal{B}$ signatures. An important tool in this regard are \emph{inclusive} searches, which are sensitive to almost all conceivable signs of baryon number violation and nicely complement the more sensitive exclusive searches.

\subsection[Exciting New Possibilities for Baryon Number Violation \\ \small\textit {Sudhakantha Girmohanta}]{\href{https://indico.fnal.gov/event/44472/contributions/193309/}{\href{https://indico.fnal.gov/event/44472/contributions/192931/}{Exciting New Possibilities for Baryon Number Violation}}}
\chapterauthor{Sudhakantha Girmohanta \\ Stony Brook University \\ E-mail: \href{mailto:sudhakantha.girmohanta@stonybrook.edu}{sudhakantha.girmohanta@stonybrook.edu}}
%\subsection{Sudhakantha Girmohanta (\href{mailto:sudhakantha.girmohanta@stonybrook.edu}{sudhakantha.girmohanta@stonybrook.edu})\\
%\href{https://indico.fnal.gov/event/44472/contributions/192931/attachments/132499/162885/ACFI20_Girmohanta.pdf}{Exciting New Possibilities for Baryon Number Violation}}

Baryon number, denoted $\mathcal{B}$, is an accidental global symmetry in the
Standard Model (SM) and is expected to be violated in many ultraviolet
extensions of it. For instance, grand unified theories (GUTs)
naturally violate $\mathcal{B}$, as quarks and (anti)leptons are placed in the
same representation(s) of the GUT gauge group. Baryon number violation
(BNV) is also a necessary condition for explaining the observed baryon
asymmetry in the universe \cite{Sakharov:1967dj}. These provided the impetus
for many dedicated experimental searches looking for baryon-number violating nucleon
decays since the early 1980s. Neutron-antineutron ($n\rightarrow\bar{n}$)
oscillations are another kind of BNV that violates $\mathcal{B}$ and $\mathcal{B-L}$ by
two units, where $\mathcal{L}$ is the total lepton number. From a low-energy
effective field theory point of view, the effective Hamiltonian ${\cal
  H}_{eff}^{(n \bar n)}$ that mediates $n\rightarrow\bar{n}$ oscillations
involves six-quark operators and therefore has coefficients with
free-field mass dimension $-5$. On the other hand, single nucleon
decay can be mediated by four-fermion operators having coefficients
with mass dimension $-2$. Thus naively one would expect $n\rightarrow\bar{n}$
oscillations to be highly suppressed when compared with single-nucleon
decay modes, but the assumption of a single mass scale responsible for
BNV might be oversimplified \cite{Mohapatra:1980qe, Rao:1982gt, Rao:1983sd}, and the
existence of new scales might suppress nucleon decay while mediating
$n\rightarrow\bar{n}$ oscillations at a level comparable to current
experimental limits. Nussinov and Shrock demonstrated this interesting
possibility in a large extra-dimensional model where fermions have
strong localization in the extra-dimensions \cite{Nussinov:2001rb}.

Let us consider a model in the context of large extra dimensions, with
the property that SM fields can propagate in the $n$ compact extra
spatial dimensions and the zero-mode fermion solutions have strong
localization at various points in the extra-dimensional space with
Gaussian profiles \cite{ArkaniHamed:1999dc, Mirabelli:1999ks}. One field-theoretic mechanism of
obtaining this localization is to couple SM fermions with appropriate
kink ($n=1$) or vortex ($n=2$) solutions. SM fermions are restricted
to an interval of length $\mathcal{L}$ in the extra dimensions. We choose the
value $\Lambda_L \equiv L^{-1} \simeq 100 \ {\rm TeV}$, i.e., $L
\simeq 2 \times 10^{-19} \ {\rm cm}$, to be consistent with bounds
from precision electroweak constraints, collider searches, and
flavor-changing neutral current processes \cite{Delgado:1999sv}. We note
that this type of model is different from extra-dimension models in
which only gravitons can propagate in the extra dimensions, as the
compactification length is much larger there \cite{Antoniadis:1998ig,Antoniadis:1990ew, Dienes:1998vh}. We use a low-energy effective field-theoretic
approach where one integrates over the short-distance physics
associated with the extra-dimensions to obtain the effective
Lagrangian in the 4-dimensional long-distance theory. This results in
a strong exponential suppression for the coefficient of an operator
that involves fermions which are localized far away from each other in
the extra-dimensional space. Therefore, this framework has the
attractive feature that it can explain the observed fermion mass
hierarchy in the SM from the locations of the corresponding chiral
parts of the fermion wavefunctions in the extra-dimensional
space. Furthermore, BNV nucleon decays can be exponentially suppressed
to be safely small by arranging sufficient separation of quark and
lepton wavefunctions from each other in the extra dimensions. This,
however, does not suppress $n\rightarrow\bar{n}$ oscillations, as the effective
Hamiltonian ${\cal H}_{eff}^{( n \bar n)}$ only involves quark fields;
indeed, $n\rightarrow\bar{n}$ oscillations may occur at
a rate comparable to current experimental limits \cite{Nussinov:2001rb}. 
Various nucleon and dinucleon decays to leptonic final
states generated by a local operator are also suppressed beyond the
current experimental sensitivity \cite{Girmohanta:2019fsx,Girmohanta:2019cjm}.

It is valuable to investigate the physics of large extra dimensions in
an enlarged gauge group, namely the left-right symmetric (LRS) group
$G_{LRS} \equiv SU(3)_c \otimes SU(2)_L \otimes SU(2)_R \otimes
U(1)_{\mathcal{B-L}}$ \cite{Mohapatra:1974gc, Senjanovic:1975rk, Mohapatra:1980yp}. Here $\mathcal{B-L}$ is gauged and is
broken when the triplet Higgs field $\Delta_R$ transforming according
to the $(1,1,3)_2$ representation of $G_{LRS}$ obtains a vacuum
expectation value (VEV) $v_R$, thereby breaking $\mathcal{B-L}$ by two
units. For a process having $\Delta \mathcal{L} = 0$ this implies
$|\Delta \mathcal{B}| = 2$.
Hence the effective scale mediating $n\rightarrow\bar{n}$ oscillations is
$v_R$ in this model. We show that, similar to the SM large
extra-dimensional framework, nucleon and dinucleon decays to leptonic
final states are exponentially suppressed while $n\rightarrow\bar{n}$
oscillations are not \cite{Girmohanta:2020qfd, Girmohanta:2020eav}. This effect is even more
enhanced in the LRS model with large extra dimensions. This is because
some operators contain only one type of fermion field, namely the
right-handed quark doublet $Q_R$. Thus, there are no multiple fields to
separate from each other in the extra dimensions, and the integration
over these extra dimensions yields no exponential suppression for this
class of operators. From experimental bounds on $n\rightarrow\bar{n}$
oscillations, this yields a bound on the effective scale $v_R$
which is substantially higher than its SM effective scale counterpart.

In a large extra-dimensional model with localized fermions it is easy
to suppress nucleon and dinucleon decays to leptonic final states by
separating out quarks and leptons from each other in the extra
dimensions. The $n\rightarrow\bar{n}$ oscillations are special in the context of
these models as they only involve quark fields and separating quarks
and leptons from each other does not suppress $n\rightarrow\bar{n}$
oscillations, which may occur at a rate comparable to current
experimental limits. In this case, $n\rightarrow\bar{n}$ oscillations and the
associated dinucleon decays become the main manifestation of BNV. This
effect is even more enhanced for the left-right symmetric model with
large extra dimensions. These findings motivate further
experimental searches for $n\rightarrow\bar{n}$ oscillations at the European
Spallation Source (ESS) \cite{Phillips:2014fgb} and the resulting matter instability in
Hyper-Kamiokande and DUNE \cite{Abi:2020evt, Barrow:2019viz}.

\subsection[Perspectives on Baryon Number Violation\\ \small\textit {David McKeen}]{\href{https://indico.fnal.gov/event/44472/contributions/193738/}{Perspectives on Baryon Number Violation}}
\chapterauthor{David McKeen \\ TRIUMF \\ E-mail: \href{mailto:mckeen@triumf.ca}{mckeen@triumf.ca}}
%\subsection{David McKeen (\href{mailto:mckeen@triumf.ca}{mckeen@triumf.ca})\\
%\href{https://indico.fnal.gov/event/44472/contributions/193738/attachments/132507/162897/Snowmass.pdf}{Perspectives on Baryon Number Violation}}

The approximate conservation of baryon number can explain the empirical fact that the proton does not appear to decay. However, we know that baryon number conservation is violated in the Standard Model at finite temperature. Moreover, the asymmetry of matter over antimatter in our Universe requires its violation. Proton decay searches can directly probe high scales where baryon number is violated by one unit around $10^{16}~\rm GeV$. In contrast, neutron-antineutron oscillation searches currently probe baryon number violation by two units at the $10^{5}~\rm GeV$ scale. Despite the much lower scales that we can currently access experimentally, $|\Delta \mathcal{B}|=2$ processes can be directly tied to the baryon asymmetry of the Universe, particularly in scenarios where the asymmetry is generated at relatively low temperatures~\cite{Babu:2013yca,McKeen:2015cuz,Allahverdi:2017edd,Aitken:2017wie,BhupalPresentation}. Such scenarios can ameliorate some cosmological problems faced in other setups. On the nodel building side, theories that admit low scale $|\Delta \mathcal{B}|=2$ processes without proton decay can be simply constructed~\cite{Arnold:2012sd}. 

For all of these reasons, experimental efforts to study $|\Delta \mathcal{B}=2|$ observables such as $n\rightarrow\bar{n}$ oscillations and dinucleon decay are vital. Additionally, efforts can be undertaken to explore other novel phenomenology that can occur in theories of this type, such as exotic hadron~\cite{Hambye:2017qix,Heeck:2019kgr} decay or the decay of atomic hydrogen~\cite{McKeen:2020zni}.

\section{Experimental Overviews}

\subsection[The European Spallation Source and Future Free Neutron Oscillation Searches\\ \small\textit {Valentina Santoro}]{\href{https://indico.fnal.gov/event/44472/contributions/192040/}{The European Spallation Source and Future Free Neutron Oscillation Searches}}
\chapterauthor{Valentina Santoro \\ European Spallation Source \\ E-mail: \href{mailto:Valentina.Santoro@ess.eu}{Valentina.Santoro@ess.eu}}
%\subsection{Valentina Santoro (\href{mailto:Valentina.Santoro@ess.eu}{Valentina.Santoro@ess.eu})\\
%\href{https://indico.fnal.gov/event/44472/contributions/192040/attachments/132318/162529/santoro.pdf}{The European Spallation Source and Future Free Neutron Oscillations Searches}}
The European Spallation Source ESS, presently under construction in Lund, Sweden, is a multi-disciplinary international laboratory. It will operate the world's most powerful pulsed neutron source. Taking advantage of the unique potential of the ESS, the NNBAR collaboration has proposed a two-stage program of experiments to perform high precision searches for neutron conversions in a range of baryon number violation (BNV) channels culminating in an ultimate sensitivity increase for $n\rightarrow\bar{n}$ oscillations of three orders of magnitude over the previously attained limit obtained at the Institut Laue-Langevin ILL\citep{BaldoCeolin:1994jz}.

\begin{figure}[h]
\centering
\includegraphics[width=0.9\textwidth]{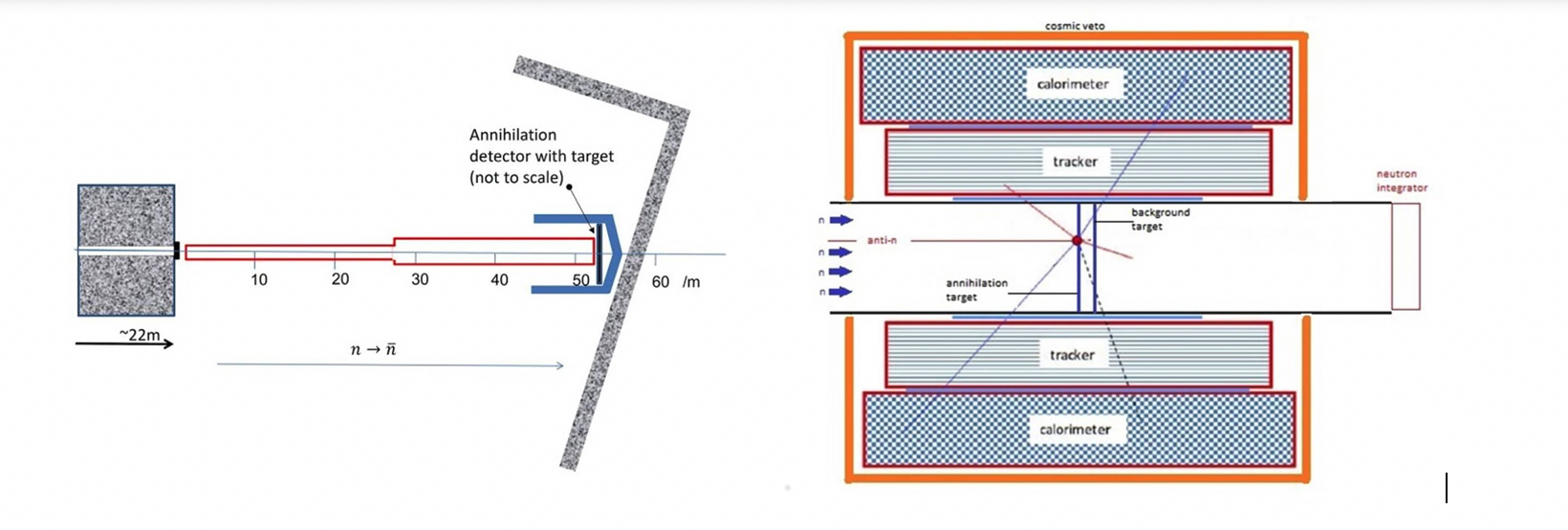}\vspace{0.6cm}
\caption{HIBEAM beamline (left), annihilation detector (right)  }
\label{fig:Santoro-Fig2}
\end{figure}

The first stage of this program, HIBEAM (the High Intensity Baryon Extraction and Measurement beamline) \citep{Addazi:2020nlz}, will employ the fundamental physics beamline during the first phase of the ESS operation, and can create a meaningful experimental program before ESS reaches its full operation power. This stage focuses principally on searches for neutron conversion to sterile neutrons $n'$. Shown in Fig.~\ref{fig:Santoro-Fig2}, one can see the HIBEAM configuration for an $n\rightarrow\bar{n}$ run. Neutrons would be transported  $\sim50\,$m through a magnetically shielded region. Any neutrons which convert into antineutrons would annihilate on a carbon foil target surrounded by an annihilation detector, which would then observe a final state comprised of $\sim 4$-$5$ pions \citep{Golubeva:2018mrz,Barrow:2019viz}. Fig.~\ref{fig:Santoro-Fig2} also shows the annihilation detector. A typical particle physics detector must be built with a tracker and a calorimeter to measure the mesonic final state after an annihilation; a cosmic veto shield is planned to suppress background. 

\begin{figure}[h]
\centering
\includegraphics[width=0.9\textwidth]{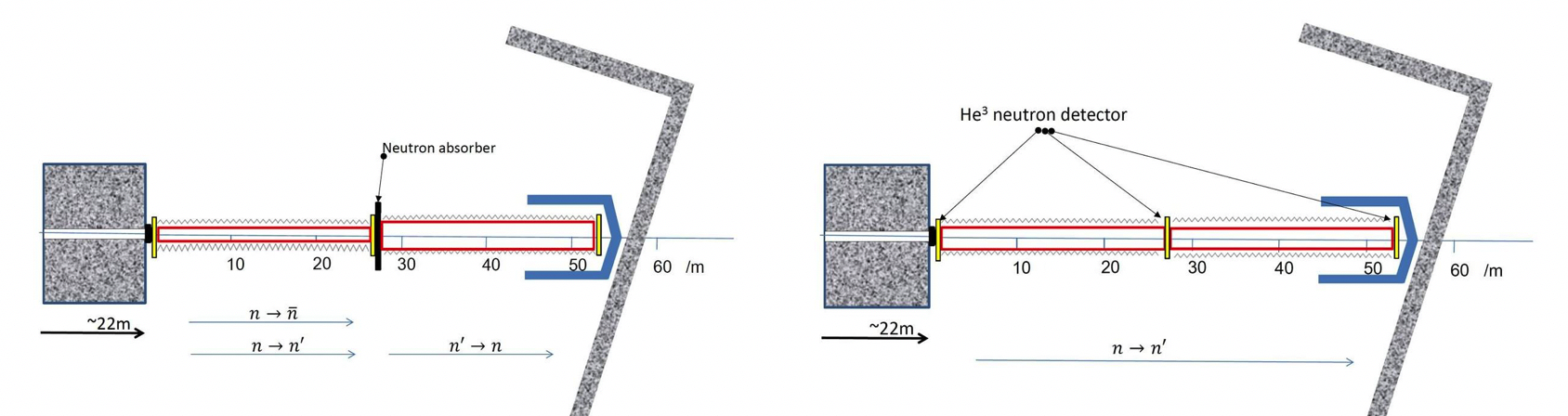}\vspace{0.6cm}
\caption{HIBEAM beamline for sterile neutron searches by regeneration (left, $n\rightarrow n' \rightarrow n$) and disappearance (right, $n\rightarrow n'$)}
\label{fig:Santoro-Fig3}
\end{figure}

Fig.~\ref{fig:Santoro-Fig3}~shows HIBEAM in a $n\rightarrow n'$ operating mode. In the regeneration mode ($n\rightarrow n' \rightarrow n$), ordinary neutrons are stopped in a total beam absorber. However, any neutrons which convert into their sterile (dark) neutron cousins \citep{Berezhiani:2006je} can pass through and convert back to neutrons, which are in turn measured by a neutron counter after 50m of propagation length. A disappearance mode ($n\rightarrow n'$) is also shown in which the flux of neutrons is measured as a function of propagation length, wherein sterile neutrons would give rise to anomalous losses of flux.\\ 

As the second stage of this compelling program at ESS, NNBAR will exploit the Large Beam Port (LBP), a unique future of the ESS facility. This beamline, in fact, begin \textit{in} the  monolith, a critical provision made for the NNBAR experiment. A normal beamport would be too small for NNBAR to reach its ambitious sensitivity goals, and so angular acceptance has been prioritized. Therefore, part of the beam extraction system has been engineered so that a large frame covering the size of three standard beamports has been constructed. Initially, the frame will be filled by three regular-size beamports which can later be removed to provide the LBP to NNBAR throughout the planned three year duration of the experiment, and eventually replaced at its conclusion.

The 2020 Update for the European Strategy for Particle Physics explicitly highlights the need for  programs at the so-called \textit{intensity frontier} at other European laboratories together with the energy frontier research at CERN. In this context, a fundamental physics program at the ESS, with a  series of measurements and searches with unique potential and sensitivity represents a compelling possibility which should not be missed.

\subsection[Neutron-antineutron oscillation search at Super-Kamiokande\\ \small\textit {Linyan Wan}]{\href{https://indico.fnal.gov/event/44472/contributions/193124/}{Neutron-antineutron oscillation search at Super-Kamiokande}}
\chapterauthor{Linyan Wan \\ Boston University \\ E-mail: \href{mailto:wanly@bu.edu}{wanly@bu.edu}}
%\subsection{Linyan Wan (\href{mailto:wanly@bu.edu}{wanly@bu.edu})\\
%\href{https://indico.fnal.gov/event/44472/contributions/193124/attachments/132299/162577/NNbar_talk.pdf}{Neutron-antineutron oscillation search at Super-Kamiokande}}
As a baryon number violating process with $\Delta \mathcal{B} = \Delta \mathcal{(B-L)} = 2$, neutron-antineutron oscillation ($n\rightarrow\bar{n}$) provides an important candidate and a unique probe to the baryon asymmetry. 
We performed a search for $n\rightarrow\bar{n}$ with the Super-Kamiokande (SK) experiment with SK-I-IV data set, corresponding to 6050.3 days of live-time. 
From last public result of SK-I~\cite{Abe:2011ky}, we updated the data set, hadron production and final state interaction model, and employed a multi-variate analysis (MVA) to better separate background and signal.
Compared to atmospheric neutrino backgrounds, due to the mechanism of $\bar nn$ or $\bar np$ annihilation within oxygen nucleus, $n\rightarrow \bar n$ signal events are generally more kinetically constrained, have more rings, and the rings are more isotropically distributed.
The MVA algorithm was derived with 12 variables describing these features and optimized towards the best sensitivity, where the total signal efficiency is $4.1\%$.
The systematic uncertainties in this search was estimated at 33\% for signal efficiency and 28\% for background rate, dominated by physics simulation such as hadronization and final state interaction as well as detector responses and reconstructions.
We observed 11 events from data, compared with the expected number of background events $9.3\pm3.0\text{ (stat.)}\pm2.6\text{ (sys.)}$. 
No statistically significant excess is observed, and the lower limit of neutron lifetime is calculated as $3.6\times10^{32}$ years at 90\% C.L., corresponding to a lower limit on neutron oscillation time $\tau_{n\bar n}=4.7\times10^{8}$ s.

\subsection[Search for $n\rightarrow\bar{n}$ in the Deep Underground Neutrino Experiment \\ \small\textit {Joshua L. Barrow and Yeon-jae Jwa}] {\href{https://indico.fnal.gov/event/44472/contributions/192778/}{Search for $n\rightarrow\bar{n}$ in the Deep Underground Neutrino Experiment} }
\chapterauthor{Joshua L. Barrow \\ University of Tennessee \\ E-mail: \href{mailto:jbarrow3@vols.utk.edu}{jbarrow3@vols.utk.edu} \\ \textit{and} \\ Yeon-jae Jwa \\ Columbia University}
%\subsection{Joshua L. Barrow ((\href{mailto:jbarrow3@vols.utk.edu}{jbarrow3@vols.utk.edu})\\
%\href{https://indico.fnal.gov/event/44472/contributions/192778/attachments/132281/162464/go}{Search for $n\rightarrow\bar{n}$ in the Deep Underground Neutrino Experiment}}
The Deep Underground Neutrino Experiment (DUNE) utilizes Liquid Argon Time Projection Chamber (LArTPC) technology to deeply probe $\nu$ and beyond Standard Model (BSM) interactions with impressive granularity. Though designed with long-baseline $\nu$-oscillation studies in mind, the large number of constituent nucleons offers impressive gains for matter instability searches. The DUNE Technical Design Report (TDR) \citep{Abi:2020evt} prioritizes BSM searches for baryon number violation (BNV) modes such as proton decay ($p\rightarrow K^+ \bar{\nu}$) and neutron-antineutron transformation ($n\rightarrow\bar{n}$) \citep{Abi:2020kei}.

\begin{figure}[h!]
    \centering
    \includegraphics[width=0.30\columnwidth]{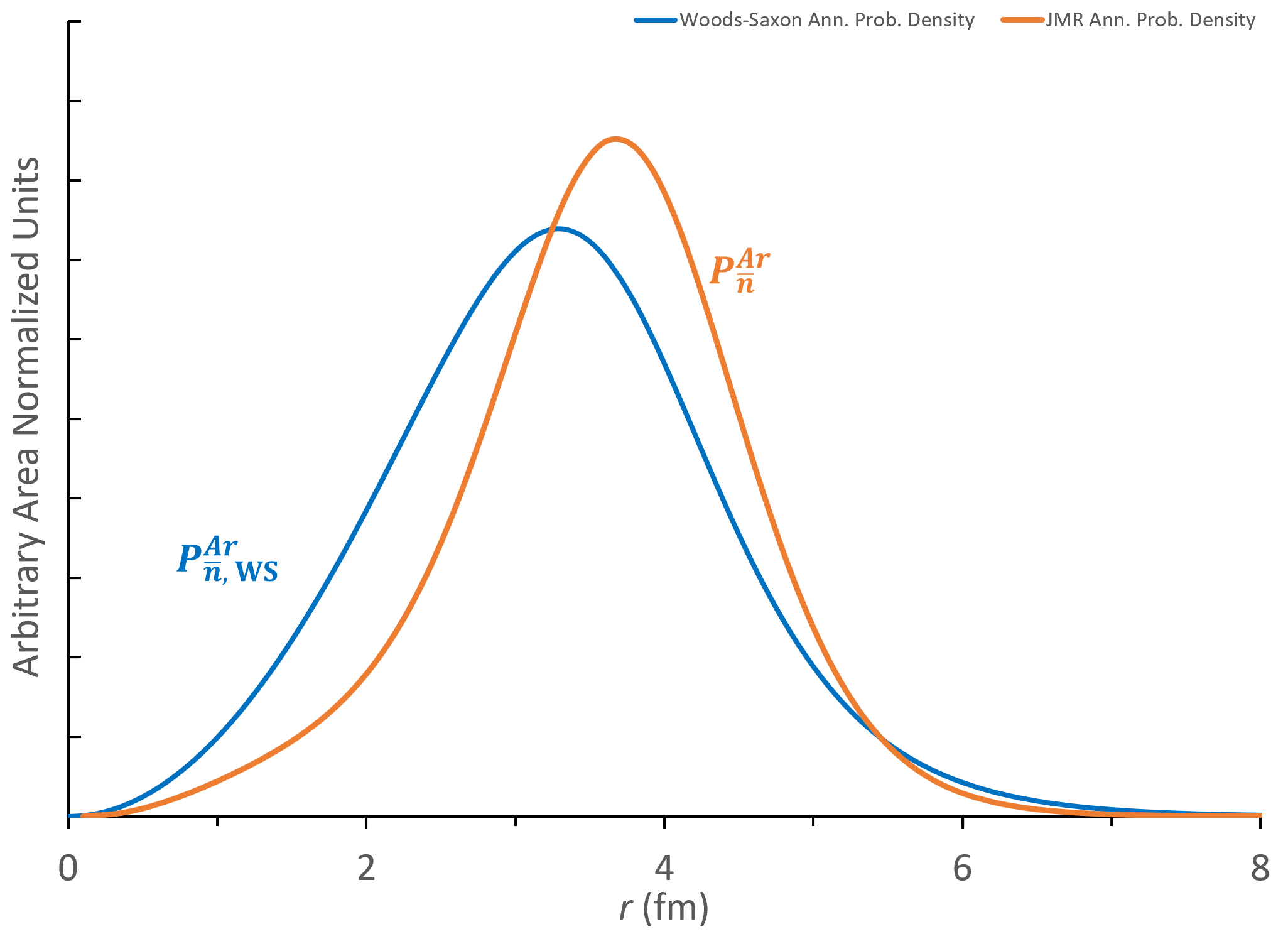}
    \includegraphics[width=0.30\columnwidth]{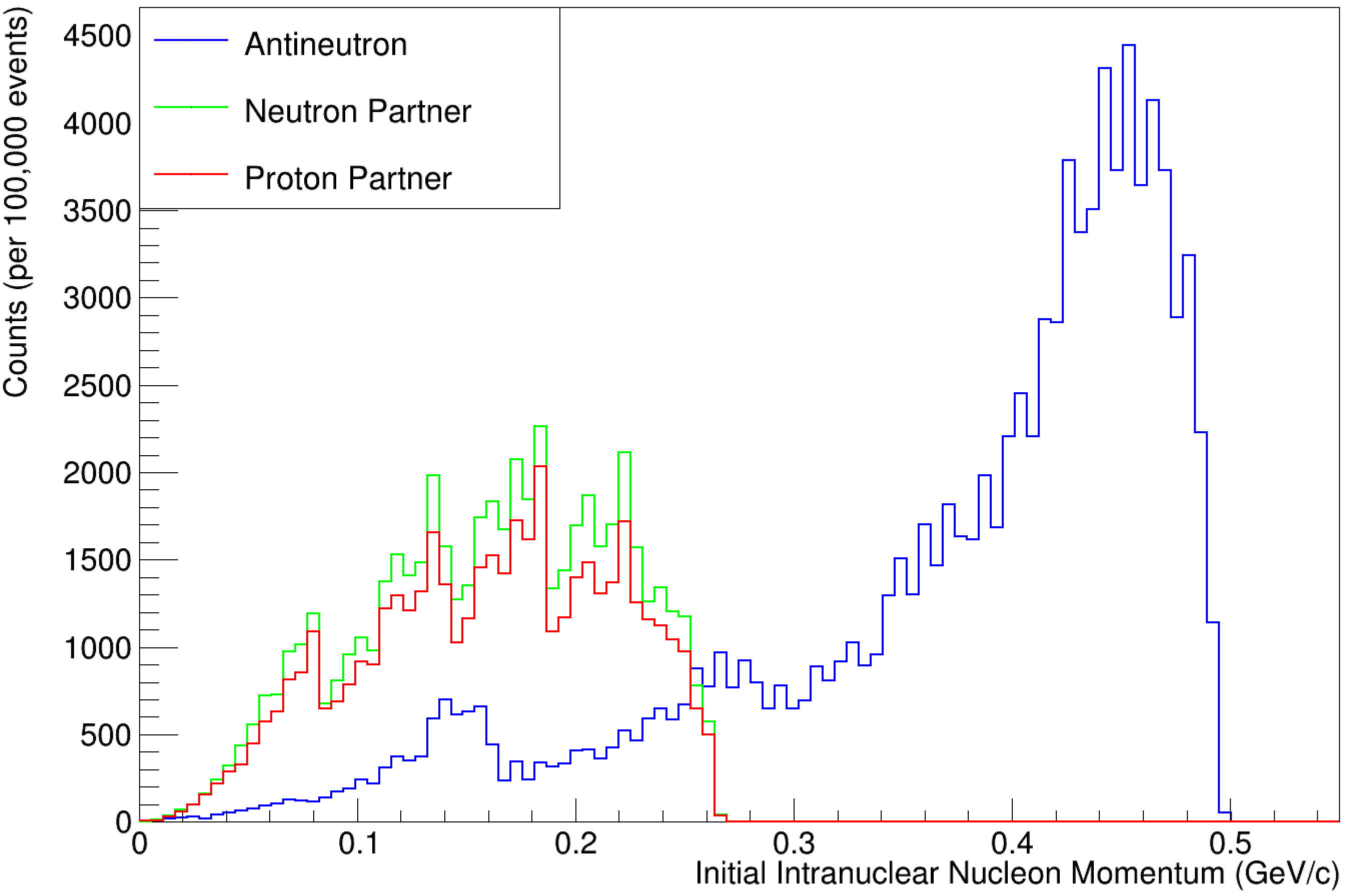}
    \includegraphics[width=0.30\columnwidth]{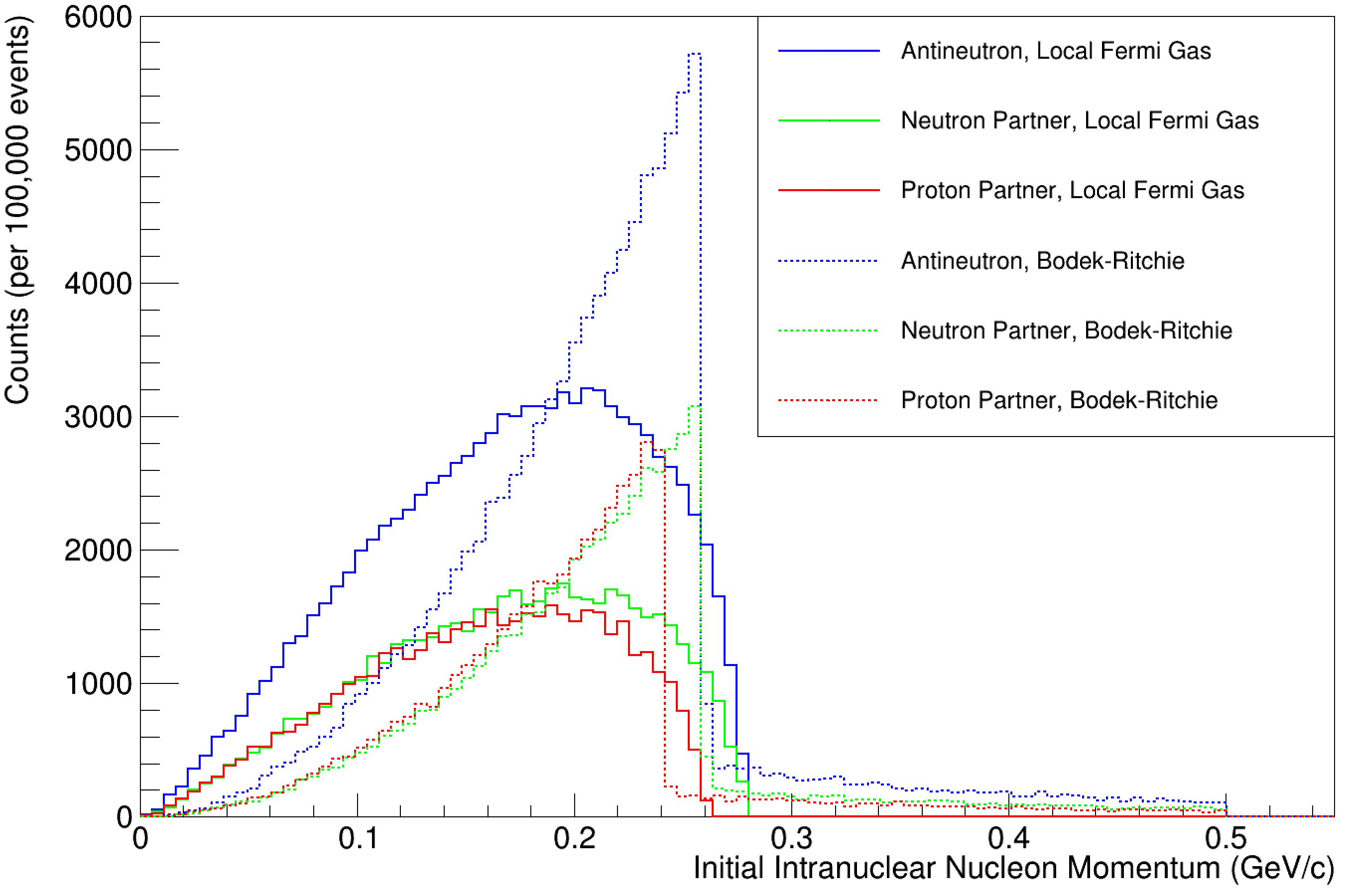}
    \includegraphics[width=0.30\columnwidth]{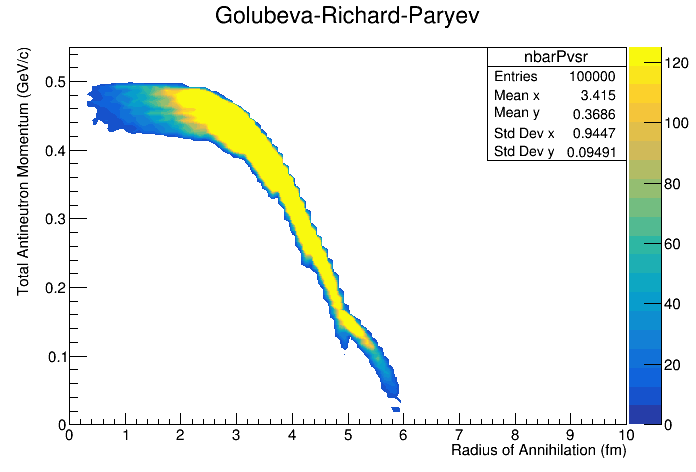}
    \includegraphics[width=0.30\columnwidth]{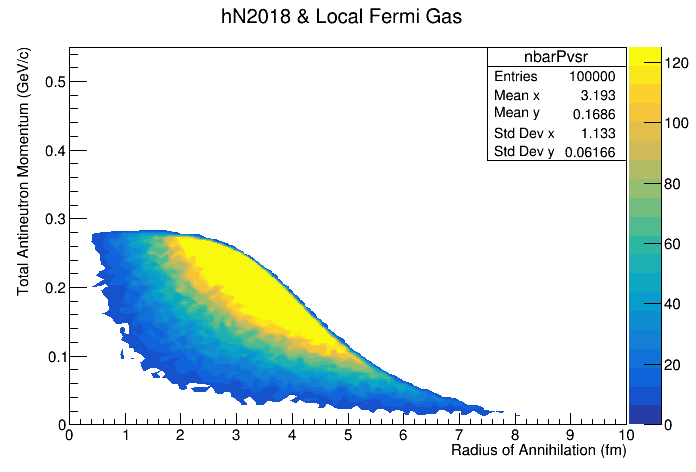}
    \includegraphics[width=0.30\columnwidth]{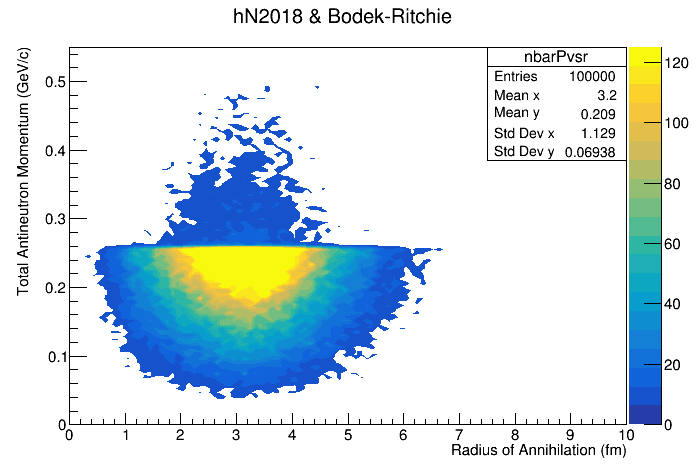}
    \caption{\textbf{Top Left}: Two curves are shown for various generator assumptions. In blue is the naive intranuclear radial position of $\bar{n}$ annihilation, a probability distribution generated by a Woods-Saxon nuclear density as presented in GENIE\citep{Andreopoulos:2009rq}. In orange is the modern, quantum-mechanically derived intranuclear radial position of annihilation probability distribution \citep{Barrow:2019viz}. The scale is arbitrary. 
    \textbf{Top Center}: The initial (anti)nucleon momentum distributions are shown using a local Fermi gas model with an additional $\bar{n}$ potential\citep{Barrow:2019viz}.
    \textbf{Top Right}: The same for the GENIEv3.0.6\citep{Andreopoulos:2009rq}, showing a local Fermi gas model and the default nonlocal Bodek-Ritchie model.
    \textbf{Bottom Left}: A two dimensional plot of intranuclear $\bar{n}$ momentum-radius correlation is shown using a local Fermi gas and the newly-derived annihilation position distribution\citep{Barrow:2019viz} (top left, orange).
    \textbf{Bottom Center}: The same using GENIEv3.0.6's\citep{Andreopoulos:2009rq} \textit{local} nonrelativistic Fermi gas nuclear model of (anti)nucleon momentum and a Woods-Saxon nuclear density (top left, blue), showing good correlation.
    \textbf{Bottom Right}: The same using GENIEv3.0.6's\citep{Andreopoulos:2009rq} \textit{nonlocal} Bodek-Ritchie relativistic Fermi gas nuclear model of (anti)nucleon momentum and a Woods-Saxon nuclear density (top left, blue), showing no positional correlation, and thus over-selecting high momenta. This parameter space shows that these last two GENIE models \textit{are not} reweightable to one another.}
    \label{fig:AnnDist}
\end{figure}

First Monte Carlo generation samples for each of these compelling BSM processes has been completed alongside an associated background of atmospheric neutrinos (which will not be discussed further here for brevity), and each has undergone full detector simulation and simulated reconstruction; details are further discussed within the TDR \citep{Abi:2020evt}. Each utilizes an optimum combined \textit{automated} approach: simulated reconstructed variables, alongside a topological differentiation score derived from a convolutional neural network (CNN), are fed into a multivariate boosted decision tree (BDT) analysis tool for hyperdimensional signal selection on an event-by-event basis. Focusing particularly on $n\rightarrow\bar{n}$, this method has shown an expected $\tau_{n\bar{n}}\geq 5.58\times10^8\,$s lower limit target is possible for DUNE \citep{Abi:2020evt,Abi:2020kei}; this is within striking distance of Super-Kamiokande \citep{WanNeutrino2020}. Considering the known capabilities of LArTPCs, this points to a need for better understanding of underlying modeling of this (these) unknown process(es) beyond default nuclear model configurations (NMCs) of Fermi motion and final state interactions (intranuclear cascades), particularly in how such automated methods respond to various disparate simulated inputs for both signal and background outputted from event generators.

\begin{figure}[h!]
    \centering
    \includegraphics[width=0.35\columnwidth]{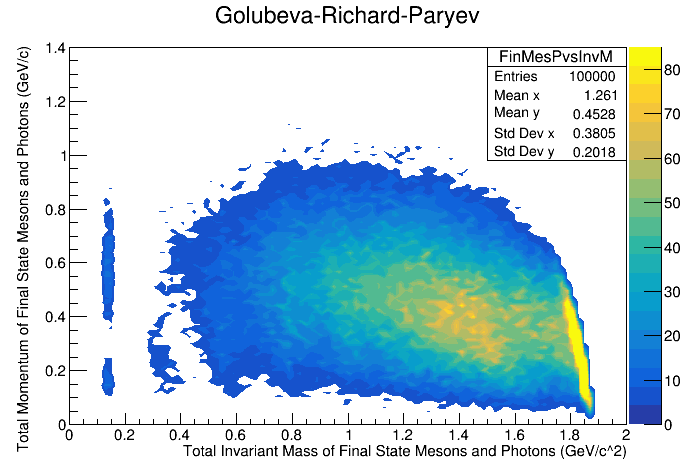}
    \includegraphics[width=0.35\columnwidth]{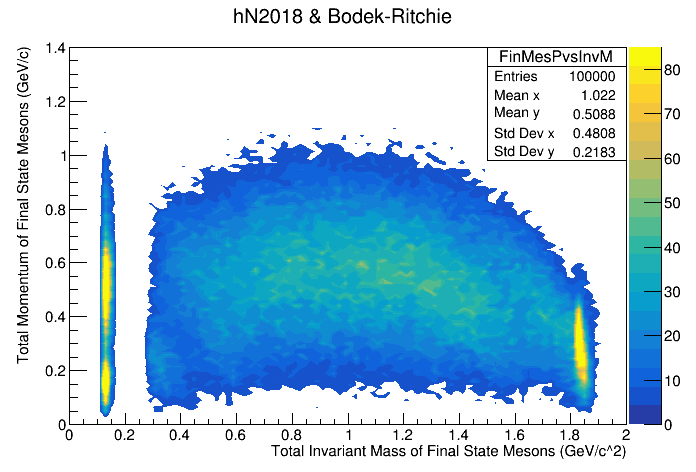}
    \caption{The final state mesonic/pionic parameter space (total momentum versus invariant mass)\citep{Abe:2011ky} after stochastic intranuclear transport of $\bar{n}$ annihilation generated mesons, compared for a few NMCs, not including detector effects. The ROI is generally considered to the be "hot-spot" in the lower right hand corner, implying the expected low Fermi momentum and high invariant mass derived from the annihilation of two nucleons creating a topologically spherical $\pi$-star; differences in these may lead to different detector signal efficiencies via automated methods.
    \textbf{Left}: a local Fermi gas model with an additional $\bar{n}$ potential and a full intranuclear cascade\citep{Golubeva:2018mrz,Barrow:2019viz}.
    \textbf{Right}: GENIEv3.0.6\cite{Andreopoulos:2009rq} using the default nonlocal Bodek-Ritchie relativistic Fermi gas and a full intranuclear cascade via the 2018 hN Intranuke model.}
    \label{fig:FinMesonTotPvInvMass}
\end{figure}

Thus, ongoing studies utilizing similar procedures are now moving toward understanding the intranuclear modeling systematics related to these unknown rare processes. By taking a ``universe"-style approach, \textit{i.e.}, iterating over various NMCs of, for instance, local and nonlocal Fermi gas models of Fermi motion (momentum), one can measure the automated method's outputs in the form of signal to background ratios via multiple pairwise comparisons across the available model space, allowing for an approximation of rare process model uncertainties beyond simple knob-turning of free parameters (such as the level of the Fermi momentum). This has been completed within an independent $n\rightarrow\bar{n}$ generator \citep{Golubeva:2018mrz,Barrow:2019viz} and GENIE \citep{Andreopoulos:2009rq} for comparisons \citep{Barrow:2019viz}, as seen in Fig. \ref{fig:AnnDist}.

By comparing these models of the initial annihilating state before nuclear transport, one can already begin to understand how various regions of experimental interest attempting to understand the final state may be affected before eventual injection into the CNN/BDT (for analysis and trigger studies). For instance, when using a local Fermi gas (Fig. \ref{fig:AnnDist}, bottom center), event generators are able to model the reduction of Fermi momentum as a function of radius, and, similarly, the lesser number of final state interactions of $\bar{n}A$ annihilation-generated mesons at higher radii; in turn, this can change the expected event topology, its expected final state reconstructed invariant mass, and total momentum.

The effects of these changes can be observed in the final state, where the region of interest, generally set around the ``hot spot" expected at high invariant mass ($\sim 2\,$GeV/$c^2$) and low total momentum, takes on different characteristic shapes and populations depending on a given NMC. The fact that these regions can be so disparate may in turn confuse automated analysis methods' responses, and so too their outputted expected signal efficiencies and background rejection rates (\textit{i.e.} their expected $\tau_{n\bar{n}}$ lower limits). Analysis is ongoing to better understand these effects as a function of the chosen NMC; comparisons will be made across signal and background in a pairwise fashion, each with their own individually trained CNNs and BDTs before intermodel comparisons. Such studies will better illuminate model's expected ranges of potential model systematics, as well as DUNE's discovery potential.

Beyond this ongoing model-focused work, there is much promise in independent automated techniques aiming to better understand particle identities from reconstructed variables in LArTPCs. Novel techniques employed by C. Sarasty (Cincinatti) and his fellow group members show great promise in their ability to differentiate proton, pion, kaon, muon, and shower candidates within LArTPCs, allowing for a precision understanding of signal or background event candidates as a whole. Such techniques could be used to improve BDT responses alongside current CNN topological differentiation techniques, allowing for a ``score" to be developed for counting of final state pions. Work to implement this within our analysis is underway, and we hope that such progress will greatly empower DUNE's BSM physics searches, including and beyond baryon number violation.

\subsection[Possible Use Of Neutron Optics for Optimization of a Free Neutron-Antineutron Oscillation Search\\ \small\textit {W. M. Snow}]{\href{https://indico.fnal.gov/event/44472/contributions/192205/}{Possible Use Of Neutron Optics for Optimization of a Free Neutron-Antineutron Oscillation Search}}
\chapterauthor{W. M. Snow \\ Indiana University \\ E-mail: \href{mailto:wsnow@indiana.edu}{wsnow@indiana.edu}}
%\subsection{W. M. Snow (\href{mailto:wsnow@indiana.edu}{wsnow@indiana.edu})\\
%\href{https://indico.fnal.gov/event/44472/contributions/192205/attachments/132292/162480/nnbarresettalkWMSAFCI.pdf}{Possible Use Of Neutron Optics for Optimization of a Free Neutron-Antineutron Oscillation Search}}
Neutron-antineutron oscillations can survive sufficiently coherent interactions with matter and external fields without suppressing the oscillation rate. I describe some examples of this phenomenon which might find practical applications in the design of future experiments. In particular, I discuss the status of neutron optics calculations which analyze what happens to an oscillating neutron-antineutron system upon reflection from a neutron mirror. Recent work \citep{Nesvizhevsky:2018tft,Nesvizhevsky:2020vwx,Gudkov:2019gro} has shown that the oscillating neutron-antineutron system can possess a sufficiently high reflectivity, low absorption of the antineutron component, and low rate of quantum decoherence to be of interest for certain implementations of a free neutron-antineutron oscillation experiment as long as the transverse phase space of the neutron beam striking the mirror is sufficiently small. Work now in preparation for publication \citep{Protasov2020} has evaluated the status of the knowledge of antineutron scattering amplitudes from present theoretical models of antineutron-nucleus and antiproton-nucleus interactions. Ongoing work \citep{Lu2020} will evaluate the dependence of the reflectivity, antineutron absorption, and quantum decoherence rate using the presently-available theory for antineutron-nucleus scattering amplitudes as a function of nucleon number and neutron transverse momentum for slow neutrons, both for single-component mirrors and bilayer mirrors, and will identify promising analogue systems which could verify these calculations using polarized neutron reflectometry from mirrors with a strong absorption for one of the two neutron spin states.
%%contribution complete

\subsection[Search for Neutron-Antineutron Oscillations with UCN \\ \small\textit {Alexey Fomin}]{\href{https://indico.fnal.gov/event/44472/contributions/192932/}{Search for NNbar with UCN}}
\chapterauthor{Alexey Fomin, Anatolii Serebrov, Mikhail Chaikovskii, Oleg Zherebtsov, and Aleksandr Murashkin \\ National Research Center ``Kurchatov Institute" - Petersburg Nuclear Physics Institute \\ E-mail: \href{mailto:fomin_ak@pnpi.nrcki.ru}{fomin\_ak@pnpi.nrcki.ru} \\ \textit{and} \\ Elena Golubeva \\  Institute for Nuclear Research, Russian Academy of Sciences}
%\subsection{Alexey Fomin (\href{mailto:fomin_ak@pnpi.nrcki.ru}{fomin\_ak@pnpi.nrcki.ru})\\
%\href{https://indico.fnal.gov/event/44472/contributions/192932/attachments/132444/162794/Fomin.pdf}{Search for NNbar with UCN}}

\begin{figure}[t]
\centering
\includegraphics[width=0.7\textwidth]{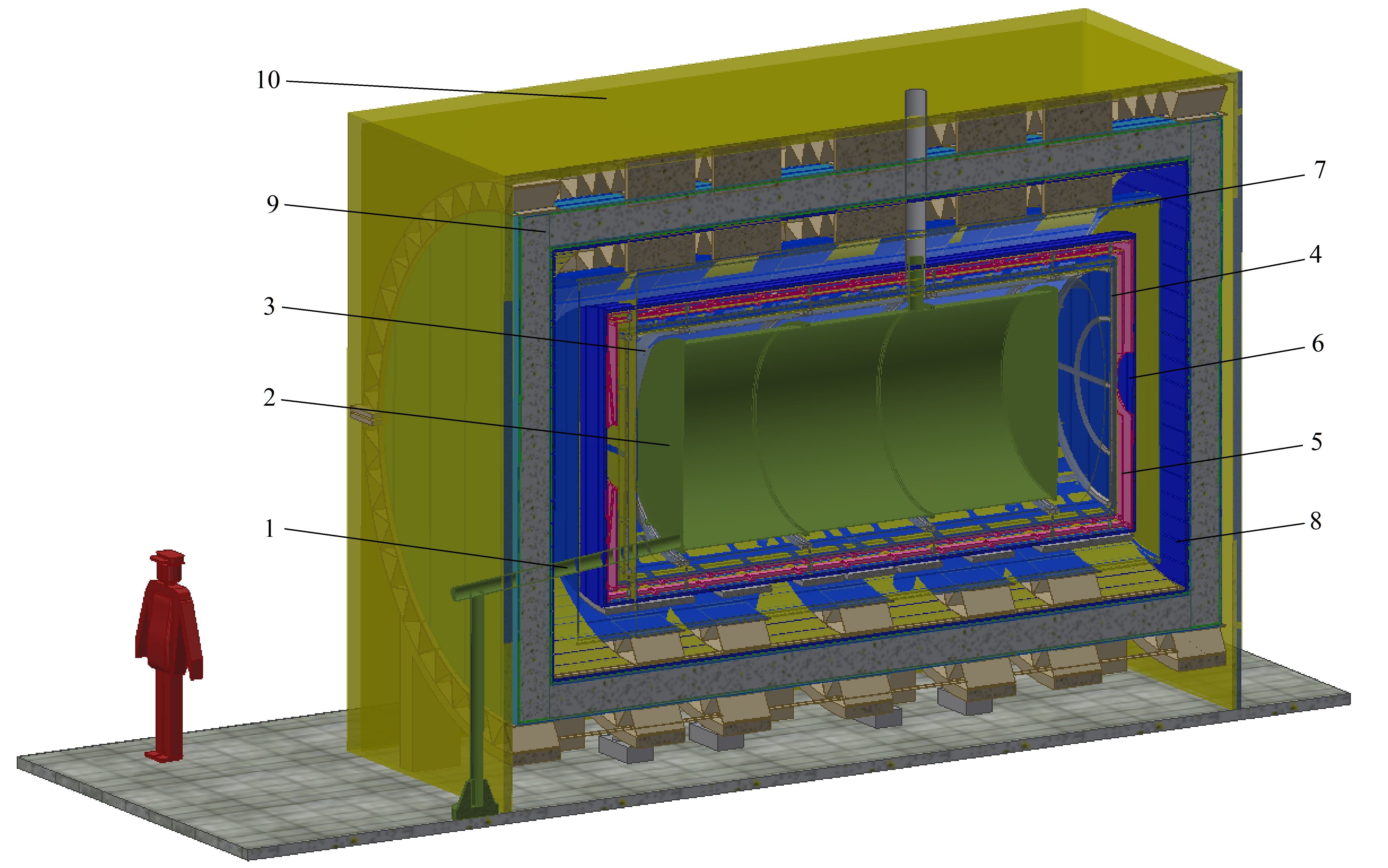}\vspace{0.6cm}
\caption{ Scheme of experimental setup: 1 – neutron guide, 2 - UCN trap, 3 - vacuum chamber, 4 – trek detector (inner part), 5 - magnetic shield, 6 - hodoscope (internal part), 7 - trek detector (middle part), 8 - hodoscope (external part), 9 - trek detector (external part), 10 – active shielding.}
\label{fig:Fomin-fig1}
\end{figure}

The scheme of the experiment for searching for neutron-antineutron oscillations based on the storage of ultracold neutrons (UCN) in a material trap is presented here; see Fig.~\ref{fig:Fomin-fig1}. In this experiment, the walls of the UCN trap play the role of the annihilation target. The idea of such an experiment has become ever more important due to the development of modern, powerful UCN sources. A prospective sensitivity of the experiment was obtained in Monte Carlo simulation modeling UCN transport and storage, mostly depending on the trap size and the amount of UCN available therein. The design of the setup, a magnetic shielding study, along with neutron storage and annihilation detection simulations were presented. The possibility of increasing the expected sensitivity of the experiment via to the accumulation of an antineutron phase \citep{Gudkov:2019gro} in the collisions of neutrons with the walls are considered. We used two models of neutron reflection from the UCN trap walls: one with partial accumulation of the antineutron phase, and another without it. The real parts of the reflection potential are close to or coinciding with the first case. For the second case, the real part of the reflection potential for antineutron is close to zero. In the first case, one expects antineutrons to reflect from walls and thus the antineutron phase is accumulated in contrast to the second case, in which no such accumulation can take place because the antineutrons immediately annihilate upon entering the wall material. However, the coefficient of antineutron reflection in the first case cannot be sufficiently high because of a large imaginary part for the reflection potential for antineutrons due to a large annihilation cross section. We utilized a UCN trap geometry in the form of a horizontal cylinder with diameter of $2\,$m and length of $4\,$m. Calculations show that the UCN source production reaches $10^8n$/s, and so the expected UCN experiment's sensitivity can be increased by about $10$–$40\times$ compared to the sensitivity of ILL experiment \citep{BaldoCeolin:1994jz}, depending on the model of neutron reflection from walls \citep{Serebrov:2016rvi,Fomin:2017aiz,Fomin:2017lej,Fomin:2018qrq,Fomin:2019oje,Fomin:2019oyj}. To calculate the expected efficiency of described detector, a GEANT4 model of the setup was created. The particles emitted from simulation-generated annihilation events \citep{Golubeva:2018mrz,Barrow:2019viz} were propogated through the setup. We studied the signal in the detector, and characteristic timescales of the events. The detector efficiency is calculated to be (68$\pm$2)\%.

\subsection[Search for neutron oscillations to a sterile state ($ n \rightarrow n' $)  and to an antineutron ($ n \rightarrow \overline{n} $) \\ \small\textit {Yuri Kamyshkov}]{\href{https://indico.fnal.gov/event/44472/contributions/192206/}{Search for neutron oscillations to a sterile state ($ n \rightarrow n' $)  and to an antineutron ($ n \rightarrow \overline{n} $)}}
\chapterauthor{Yuri Kamyshkov \\ The University of Tennessee \\ E-mail: \href{mailto:kamyshkov@utk.edu}{kamyshkov@utk.edu}}
%\subsection{Yuri Kamyshkov (\href{mailto:kamyshkov@utk.edu}{kamyshkov@utk.edu})\\
%\href{https://indico.fnal.gov/event/44472/contributions/192206/attachments/132468/162851/YK-ACFI-talk.pdf}{Search for neutron oscillations to a sterile state ($ n \rightarrow n' $)  and to an antineutron ($ n \rightarrow \overline{n} $)}}
As follows from theoretical conjectures of Z. Berezhiani et al. [2006-2020] the neutron that is part of the Standard Model ($SM$) can oscillate into sterile state $ n \rightarrow n' $ , thus leading to neutron disappearance or baryon number violation $ \Delta \mathcal{B}=-1 $. However, this can be only an apparent disappearance: if the sterile neutron $ n' $ is a part of the Mirror Standard Model ($ SM' $) with corresponding mirror baryon number $ \mathcal{B}' $ the transformation $ n \rightarrow n' $ can occur without violation of the global baryon number, i.e. with $ \Delta \mathcal{(B+B')} =0 $. This process will be not necessarily suppressed by high mass scale and can have observable probability corresponding to oscillation times as small as 1-100 s. The $ SM' $ sector is assumed to be an exact copy of $SM$ with the same particle content and the same gauge interactions within $ SM' $. But these interactions are absent between $SM$ and $ SM' $ particles, e.g. mirror photon $ \gamma' $  will not interact with $SM$ charges and vice versa. The gravity however is a common interaction for both sectors thus making $ SM' $ a good candidate for the Dark Matter. Also, additional new BSM interactions conjectured by Z. Berezhiani \citep{Berezhiani:2006je,Berezhiani:2018qqw,Berezhiani:2018zvs,Berezhiani:2020nzn} will mix the neutral particles of $SM$ and $ SM' $ sectors (the particles like $ \gamma , \nu , n $ and possibly other neutral particles) that makes such interactions responsible for the direct detection of the DM and for the transformations like $ \gamma \rightarrow \gamma',~ \nu \rightarrow \nu '~ $, and particularly interesting $ n \rightarrow n' $. The latter process will be most convenient and easy for experimental observations.

Existing neutron sources provide cold neutron beams with high intensities that can be used for corresponding experimental searches like $ n \rightarrow n' $ disappearance, $ n \rightarrow n' \rightarrow n $ regeneration, searches for the neutron transition magnetic moment, and neutron–antineutron transformations through mirror-state oscillations $ n \rightarrow n' \rightarrow \overline{n} $. The latter process should be searched in the presence of some magnetic field $B$ that will enhance transformation in a resonant way when this field $B$ will coincide with unknown value of mirror magnetic field $ B' $ that should be found by the magnetic field scan. Theoretical expectations do not exclude that the transformation effect will be large in this case. Plans for such measurements with existing neutron sources at the Oak Ridge National Laboratory and at the future European Spallation Source with the estimates of the sensitivity reach are discussed in the workshop presentation.
%contribution is complete

\subsection[Neutrons at ORNL and ESS: A Synergistic Program \\ \small\textit {Marcel Demarteau}]{\href{https://indico.fnal.gov/event/44472/contributions/192072/}{Neutrons at ORNL and ESS: A Synergistic Program}}
\chapterauthor{Marcel Demarteau \\ Oak Ridge National Laboratory \\ E-mail: \href{mailto:demarteau@ornl.gov}{demarteau@ornl.gov}}
%\subsection{Marcel Demarteau (\href{mailto:demarteau@ornl.gov}{demarteau@ornl.gov})\\
%\href{https://indico.fnal.gov/event/44472/contributions/192072/attachments/132460/162840/demarteau_nnbar.pdf}{Neutrons at ORNL and ESS: A Synergistic Program}}
High Energy Physics finds itself at a most interesting time exploring energy and matter and its evolution at its deepest level. There is more today that we do not understand about the universe than a couple of decades ago. The community is bubbling with creative new ideas that have the potential to drive a new and profound tool-driven revolution that, if history is our guide, will discover entirely new phenomena that need to be explained. Many ideas are being entertained for new projects. Currently, however, the field finds itself dominated by mega-projects that leave little room for a broad spectrum of experimental research. The scientific merit of these large projects is unquestioned and endorsed by long-range planning studies. The field of particle physics, however, stands to gain tremendously by exploiting non-traditional high-energy facilities to complement and expand its research portfolio. For example, since 2013 Oak Ridge National Laboratory has been developing the utilization of its neutron user facilities for fundamental neutrino science. Two major experiments, PROSPECT at the High Flux Isotope Reactor and COHERENT at the Spallation Neutron Source, have demonstrated that these facilities can deliver world-class neutrino science, while maintaining their commitments to their primary missions for the Office of Basic Energy Sciences. 

An inclusive approach both to the science program and to the development of facilities will allow for significant benefits for the high energy physics community. Several non-traditional HEP projects could provide unique and important contributions for the studies of particle physics looking for physics beyond the standard model and studying fundamental symmetries. The proton power upgrade at ORNL, delivering a 2.8~MW proton driver in 2025 for neutron scattering experiments, followed by the completion of the Second Target Station, are ambitious projects that can provide opportunities to inform the future high energy physics research program. The High Flux Isotope Reactor (HFIR) provides a continuous well-understood source of electron antineutrinos that is being tapped already for neutrino studies. Two upgrades of HFIR are being considered with an opportunity for a larger program in support of fundamental physics experiments, a notion that is strongly supported by a recent BESAC report. %These two powerful neutron sources, and their upgrades, can be used to search for n-nbar and mirror neutron oscillations. 
Searches for free neutron oscillations at these facilities provide unique opportunities to search for symmetry breaking mechanisms, like baryon number and baryon-minus-lepton number violation, that are complementary and necessary, in combination with the proton decay studies, to obtain a complete picture of the fundamental interactions. These experiments at ORNL could lay the foundation for second generation experiments at the European Spallation Source (ESS) if the experimental conditions at the ESS enable further significant advances. 

Although the main mission of the SNS and HFIR are the production of neutrons for neutron scattering experiments, the value of these traditional neutron facilties reaches far beyond neutron scattering. As has been demonstrated, these facilities are excellent neutrino sources with exceptional characteristics, that have already provided world-class results in neutrino physics. Given their impact, the fundamental neutrino science program in operation at both the SNS and HFIR is growing. The fundamental neutron physics beamline at the SNS is also dedicated to the study of the fundamental interactions and their symmetries. These efforts can and should be expanded. A balanced program consisting of a mix of small and large projects is required for a healthy, broadband high energy physics program and the role of the traditionally neutron facilities can be greatly expanded to advance science in fundamental interactions in a timely manner, that could be the precursor for more advanced experiments at the ESS.

\subsection[Measurements of Neutron Coupling to a Mirror Sector Using Spin Precession\\ \small\textit {Albert Young}]{\href{https://indico.fnal.gov/event/44472/contributions/193591/}{Measurements of Neutron Coupling to a Mirror Sector Using Spin Precession}}
\chapterauthor{Albert Young \\ North Carolina State University \\ E-mail: \href{mailto:aryoung@ncsu.edu}{aryoung@ncsu.edu}}
%\subsection{Albert Young (\href{mailto:aryoung@ncsu.edu}{aryoung@ncsu.edu})\\
%\href{https://indico.fnal.gov/event/44472/contributions/193591/attachments/132484/162867/Neutron_Mirror_Neutron_Precession_Experiments_2020.pdf}{Measurements of Neutron Coupling to a Mirror Sector Using Spin Precession}}
Couplings between neutrons and a mirror sector can be formulated in terms of observable effects for precession-measurements (as has been pointed out by Berezhiani), bringing to bear the tools and experimental resources already in play for the measurement of static electric dipole moments.limits for measurements with the coupling strength for neutrons to mirror neutrons. Some details of measurements in an EDM-like geometry can be used to place limits on mirror couplings and provide information on the orientation and strength of a mirror magnetic field, should it exist in the mirror sector.

\mychapter{2}{Workshop Summary}
The Amherst Center for Fundamental Interactions Workshop, ``Theoretical Innovations for Future Experiments Regarding Baryon Number Violation,'' held virtually August 3$\mathrm{^{rd}}$--7$\mathrm{^{th}}$, 2020, brought to light several key opportunities and requirements to address the origin of the baryon asymmetry in the universe (BAU) by searching for neutron-antineutron transformations ($n\rightarrow\bar{n}$). 
Observation of %a %\textit{neutron-antineutron oscillation} %neutron-antineutron oscillation or transformation process 
$n\rightarrow\bar{n}$ \citep{Mohapatra:1980qe,Glashow:1979nm,Phillips:2014fgb} 
%may reveal secrets of the matter-antimatter asymmetry in the universe.  Its observation 
would be clear evidence for baryon number ($\mathcal{B}$) violation (BNV), %the only experimentally unconfirmed of the 
one of the three Sakharov conditions \citep{Sakharov:1967dj} that has yet to be experimentally confirmed, and
%three Sakharov conditions,
which together can explain the dynamical generation of the BAU. %observed matter-antimatter asymmetry of the universe. 
%While the Standard Model conserves baryon-minus-lepton number ($\mathcal{B}-\mathcal{L}$), t
To avoid ``washing out" by Standard Model (SM) sphalerons, $\mathcal{(B-L)}$-violation is a prerequisite for any pre-existing $\mathcal{B}$ or $\mathcal{L}$ asymmetry to dynamically develop and survive; the latter is the case in classic leptogenesis. With the effective impossibility of a definitive, ``on shell" test for classic leptogenesis, similar to the confirmations of the $W^{\pm}$, $Z^0$, and Higgs, other potentially observable baryogenesis alternatives become attractive to consider. 
Since $\Delta\mathcal{(B-L)}\neq 0$ for $n\rightarrow\bar{n}$ (and more generally $\Delta\mathcal{B}=2$ dinucleon decays), the fundamental physics behind $n\rightarrow\bar{n}$ may well underlie the origin of the $\mathcal{B}$-asymmetry surviving until our current epoch. This contrasts with the ephemeral $\mathcal{B}$-asymmetry generated in grand unified theories via $\Delta\mathcal{(B-L)}=0$ processes, which can be diluted
by sphaleron effects.
%which typically does not persist due to sphaleron effects.
%are typically washed out by Standard Model (SM) sphaleron effects.

Many beyond SM (BSM) theories of baryogenesis predict $n\rightarrow\bar{n}$ in an observable range. An example is the compelling \citep{NelsonINT2017} %\textit{post-sphaleron baryogenesis}
post-sphaleron baryogenesis (PSB)  model \citep{Babu:2006xc,Babu:2008rq,Babu:2013yca} where baryogenesis occurs %\textit{after}
after the electroweak phase transition, predicting an %\textit{upper limit} 
upper limit on the $n\rightarrow\bar{n}$ oscillation period $\tau_{n\bar{n}}$ which may be within reach of forthcoming experiments. More generally, ``Majorana baryogenesis" \citep{Cheung:2013hza,Baldes:2014rda,Grojean:2018fus}, effective at low energy scales, can also lead to observable $n\rightarrow\bar{n}$. These mediating Majorana fermions could be the gluinos or neutralinos of supersymmetric models with $R$-parity violation, or can be involved in neutrino ($\nu$) mass generation \citep{Dev:2015uca}. In some cases, if certain colored scalars remain light at the TeV scale \citep{Babu:2012vc}, GUT scale BNV interactions can lead to successful baryogenesis and observable $n\rightarrow\bar{n}$. It has been shown that $n\rightarrow\bar{n}$ can also result in models where baryogenesis proceeds through the related process of particle-antiparticle oscillations of heavy flavor baryons \citep{McKeen:2015cuz,Aitken:2017wie}. This possibility points towards new physics at the scale of a few TeV, and its ingredients (heavy neutral fermions and colored scalars) could be within the reach of the Large Hadron Collider (LHC).

%\subsection{Generalities in QFT/EFT}
In a low-energy effective field theory (EFT) analysis, the leading operators contributing to proton (and bound  $n$) decay are four-fermion operators, which have dimension $d=6$, and hence coefficients of the form $1/M_{Nd}^2$, where $M_{Nd}$ denotes the mass scale characterizing the physics responsible for nucleon decay. However, these operators %\textit{conserve}
conserve $\mathcal{(B-L})$, and are thus not useful for understanding the BAU. %$\mathcal{B}$ asymmetry of the universe. 
In contrast, $n\rightarrow\bar{n}$ is mediated by six-quark operators, which have $d=9$, and so have coefficients of order $1/M_{n \bar n}^5$. If $M_{Nd} \simeq M_{n \bar n}$, then one might naively conclude that nucleon decay would be more important than $n\rightarrow\bar{n}$ as a manifestation of BNV. However, there are models in which %\textit{the opposite}
the opposite is the case, where instead nucleon decay is absent or highly suppressed while $n\rightarrow\bar{n}$ remains the dominant manifestation of BNV\citep{Mohapatra:1980qe,Glashow:1979nm,Chang:1980ey,Kuo:1980ew,Cowsik:1980np,Rao:1982gt}.
%\citep{Mohapatra:1980qe,Glashow:1979nm}; other early works include \citep{Chang:1980ey,Kuo:1980ew,Cowsik:1980np,Rao:1982gt}.

It is known that $n\rightarrow\bar{n}$ can occur naturally at observable rates in a model with a left-right-symmetric gauge group $G_{LRS}={\rm SU}(3)_c \times {\rm SU}(2)_L \times {\rm SU(2)}_R \times {\rm U}(1)_{B-L}$ \citep{Mohapatra:1974gc,Senjanovic:1975rk,Mohapatra:1980qe}. Here, $\mathcal{B}$ and $\mathcal{L}$ are connected via the $\mathcal{(B-L})$ gauge generator, and the breaking of $\mathcal{L}$ leads to Majorana $\nu$'s via the seesaw mechanism. This, in turn, can lead naturally to $n\rightarrow\bar{n}$ in a quark-lepton unified theory, while proton decay is absent in minimal versions of such models.%. This gauge group can arise via the breaking of an SU(4) gauge symmetry \citep{Pati:1974yy,Mohapatra:1974gc} to ${\rm SU}(3)_c \times {\rm U}(1)_{B-L}$. After this breaking, the theory contains Higgs fields $\phi: (1,2,2)_0$, $\Delta_L: (1,3,1)_2$, and $\Delta_R: (1,1,3)_2$ (numbers are dimensions of representations  of factor groups in $G_{LRS}$ and subscripts are $\mathcal{B-L}$ charges) together with a set of color-sextet Higgs. These can mediate $n-\bar n$ transitions at an observable level, while proton decay is absent \citep{Mohapatra:1980qe}. This LRS model makes a profound connection between $\Delta B=2$ units and 
%$\Delta L=2$, and hence Majorana $\nu$ masses, via the vacuum expectation value of the $\Delta_R$ Higgs \citep{Mohapatra:1980qe,Mohapatra:1980yp}. These Majorana $\nu$ masses drive a seesaw mechanism that provides a very appealing explanation for light $\nu$ masses. 

Another class of models with $n\rightarrow\bar{n}$ are those with extra spatial dimensions, where SM fermions can retain localized wave functions within these extra dimensions \citep{Nussinov:2001rb,Girmohanta:2019fsx,Girmohanta:2020qfd}. In such models, it is trivial to suppress nucleon decays well below experimental limits by separating the wave function centers of quarks and leptons sufficiently. %However, this does not suppress 
$n\rightarrow\bar{n}$ transitions are not suppressed because the six-quark operators do not involve leptons. In these cases, $n\rightarrow\bar{n}$ oscillations can occur at rates comparable to existing experimental limits \citep{Nussinov:2001rb,Girmohanta:2019fsx,Girmohanta:2020qfd}, and there are many explicit model examples \citep{Mohapatra:1980qe,Nussinov:2001rb,Girmohanta:2019fsx} in which nucleon decay is absent or highly suppressed. Thus, $n\rightarrow\bar{n}$ would remain the primary manifestation of BNV for forseeable terrestrial experiments. Other examples of models without proton decay but with $n\rightarrow\bar{n}$ have been discussed in \citep{Rao:1983sd,Arnold:2012sd,Allahverdi:2017edd,Gardner:2018azu}. %These models provide strong theoretical motivation for new experiments to search for $n\rightarrow\bar{n}$ and their associated destabilization of matter.

The question of the origin of the BAU may be related to that of the nature of dark matter, such as %perhaps 
via a cogenesis between ordinary and dark sectors\citep{Bento:2001rc,Bento:2002sj}. %need a smoother transition
Mirror matter, a type of hypothetical dark sector constituted by cold atomic or baryonic matter originating from a %\textit{sterile}
sterile parallel SM$'$ gauge sector (a replica of our %\textit{active} 
active SM sector), is a viable dark matter candidate \citep{Berezhiani:2003xm,Berezhiani:2000gw,Berezhiani:2003wj}. Such a sector may provide another experimental portal onto $n\rightarrow\bar{n}$ physics, as well as motivate synergistic R\&D initiatives. $\Delta\mathcal{(B-L)}=1$ interactions between SM and SM$'$ sectors may be at the origin of ordinary (active) and mirror (sterile) $\nu$ mixing \citep{Akhmedov:1992hh,Berezhiani:1995am}, 
so that mirror neutrinos can be most natural candidates for sterile neutrinos \citep{Foot:1995pa,Berezhiani:1995yi}.
%maybe we should say something about SM' temperature
Another possibility is neutron--mirror neutron mixing, leading to neutron into (sterile) mirror neutron transitions ($n\rightarrow n'$) \citep{Berezhiani:2005hv,Berezhiani:2008bc}. In the early universe, such mixing can co-generate both ordinary and mirror $\mathcal{B}$ asymmetries \citep{Bento:2001rc,Berezhiani:2018zvs}, giving a common origin to the observed baryonic and dark matter fractions of the universe, $\Omega_{\rm DM}/\Omega_{\rm B} \simeq 5$ \citep{Berezhiani:2003xm,Berezhiani:2008zza}. 

In contrast to $n\rightarrow\bar{n}$,  $n\rightarrow n'$ could be a  
fast process with an oscillation period of %mere 
seconds, and thus contain rich astrophysical implications, e.g. for ultra-high energy cosmic rays \citep{Berezhiani:2006je,Berezhiani:2011da}. 
Several experimental groups have searched for these oscillations 
%In some previous searches via $n$ disappearance ($n\to n'$) 
using ultracold neutrons (UCN) \citep{Serebrov:2007gw,Serebrov:2009zz,Ban:2007tp,Altarev:2009tg}. % and regeneration ($n\to n' \to n$), where interestingly there appear to be 
Some deviations from the null-hypothesis have been reported in $n\to n'$ disappearance searches using UCN \citep{Berezhiani:2012rq,Berezhiani:2017jkn}. The phenomena of  $n\rightarrow\bar{n}$ ($\Delta\mathcal{B}=2$) and $n\rightarrow n'$ ($\Delta \mathcal{B} = 1$) can be interrelated in unified theoretical frameworks, becoming parts of one common picture \citep{Berezhiani:2015afa}. While $n-\bar n$ oscillation does not violate discrete symmetries and in particular CP, it can be violated in $n-n'$ and/or $n-\bar n'$ oscilllations \citep{Berezhiani:2018xsx,Berezhiani:2018pcp}. 
In addition, $n-n'$ transitions can be induced not only by mass mixing but also via a transitional magnetic moment (or electric dipole moment) between $n$ and $n'$ \cite{Berezhiani:2018qqw}. 
Neutron-mirror neutron oscillation effects can be detected by looking at the neutron disappearance due to $n\to n'$ transition or by regeneration $n\to n'\to n$ \citep{Berezhiani:2017azg}. New searches are planned and ongoing using ultracold \citep{Abel:2018gpz, Abel:2020kdg} and cold neutrons \citep{Broussard:2017yev,Broussard:2019tgw}.

Interestingly, both $n-n'$ and $n-\bar n'$ mixings can exist. 
This gives rise to a novel mechanism of $n\to \overline{n}$ via an $n\to n',\bar n' \to \overline{n}$ shortcut \citep{Berezhiani:2020nzn}, whose effect can be up to ten orders of magnitude larger than the one induced by direct  $n\rightarrow\bar{n}$ mixing.

%\textit{Bob, Rabi}
%\subsection{From quarks to observables}
%\textit{Jordy, Mike}\\
%\textbf{Citations to be added }
Predictions for $\tau_{n\bar{n}}$ and dinucleon decay rates start with quark-level amplitudes for $\Delta \mathcal{B}=2$ six-quark operators, which are then matched to the hadronic level by calculations combining lattice QCD and chiral effective field theory ($\chi$EFT). Depending on the quark-level operator, different hadronic operators are induced. Typically, the most important are one-body $n\rightarrow\bar{n}$ operators, giving rise to both $n\rightarrow\bar{n}$ as well as dinucleon decays at leading order in $\chi$EFT \citep{Oosterhof:2019dlo,Haidenbauer:2019fyd}. The $n\rightarrow\bar{n}$ transition matrix elements of these operators have recently been calculated in exploratory lattice QCD calculations which directly connect the low-energy $n\rightarrow\bar{n}$ oscillation period to the parameters of BSM theories of $\mathcal{(B-L})$-violation\citep{Rinaldi:2018osy,Rinaldi:2019thf}. In $\chi$EFT, $n\rightarrow\bar{n}$ is described by a Majorana $n$ mass whose coupling can be fixed by matching to lattice QCD results. The same coupling can be used to calculate the deuteron lifetime at leading order in $\chi$EFT, but at higher-order there are additional contributions from two-body operators encoding the strength of $\Delta \mathcal{B} = 2$ nuclear interactions. The presence of these relatively unexplored interactions currently gives rise to uncertainties in determinations of BNV decays of nuclei. With improvements in the hadronic and nuclear theory, this difference could instead be turned into a feature for eventually discriminating between different BSM explanations of $\mathcal{(B-L})$-violation after observing \textit{both} free and bound $n\rightarrow\bar{n}$ in experiments. Capitalizing on recent progress in lattice QCD calculations of nuclear matrix elements\citep{Detmold:2019ghl,Davoudi:2020ngi} and \textit{ab initio} nuclear theory calculations\citep{Hammer:2019poc,Gandolfi:2020pbj} which include high-order nucleon-nucleon and nucleon-antinucleon chiral interactions, %it can be envisioned that 
the lifetimes of some heavier nuclei of experimental interest, such as ${}^{16}$O, could be reliably calculated using similar EFT methods, relying on controlled approximations to the SM to compute the required nuclear matrix elements. BSM physics parameters can be related to the lifetimes of even heavier nuclei using well-known existing nuclear models \citep{Dover:1985hk,Friedman:2008es,Barrow:2019viz}, themselves offering excellent phenomenologies to be probed.
%will require additional nuclear modeling. There
%has already been great progress on this front \textbf{Jean-Marc here}.\\

%Target Rare Processes and Precision Frontier
%\section{Experimental Opportunities} %todo add refs
Future facilities will provide compelling and complementary opportunities to further explore both BNV and dark sector candidates using free %cold, very cold, and ultracold 
$n$'s alongside more traditional intranuclear searches for $n\rightarrow\bar{n}$ and dinucleon decays. Searches for free and intranuclear $n\rightarrow\bar{n}$ are both needed to determine the source(s) of BSM physics. The European Spallation Source (ESS), currently under construction, will be the world's most powerful pulsed source of cold  $n$'s. Current and future large underground detectors such as Super-Kamiokande (SK), the Deep Underground Neutrino Experiment (DUNE)\citep{Abi:2020evt}, and Hyper-Kamiokande (HK)\citep{Abe:2018uyc} offer substantial increases in mass, exposure, and reconstruction capabilities, and thus higher sensitivities to rare processes. %--though at the cost of likely irreducible atmospheric neutrino backgrounds. % I propose we be positive in the introductory paragraph, since this point is made later
Existing US-based Basic Energy Science facilities, including but not limited to the Spallation Neutron Source (SNS) and High Flux Isotope Reactor (HFIR) at ORNL \citep{Broussard:2019tgw}, %mention other facilities to show broader community? keep to baseline?
can be leveraged for research and development for complementary science on short time scales, and are also interesting possibilities to consider with their planned future upgrades. Examples include an optimized future $100\,$MW HFIR and the planned Second Target Station at the SNS.
%\subsection{Free cold neutrons} %todo add refs

%\textit{Josh, Yuri, Matt, Leah, Valentina, David}\\
The last free $n\rightarrow\bar{n}$ search using cold  $n$'s was performed in $\sim$1990 at the Institut Laue-Langevin (ILL) \citep{BaldoCeolin:1994jz}, achieving a lower limit of $\tau_{n\bar{n}}\sim10^8\,$s. In the intervening period, there has been substantial progress in both development of advanced  $n$ optics and annihilation-generated particle detection capabilities. By taking advantage of the current state of the art at future $n$ sources, an improvement in sensitivity of $\gtrsim1000\times$ILL \citep{Frost,Klinkby,Santoro:2020nke,Phillips:2014fgb} becomes possible, reaching $\tau_{n\bar{n}}\sim10^{9-10}\,$s \citep{Nesvizhevsky:2020vwx,Addazi:2020nlz,Gudkov:2019gro}. The most promising opportunity for a future free $n\rightarrow\bar{n}$ search comes from an ambitious proposal by the NNBAR Collaboration \citep{Phillips:2014fgb,Addazi:2020nlz} at the ESS. The ESS has %recognized the importance of fundamental physics in its program\cite{something}, and 
included an important design accommodation for NNBAR to achieve the high $n$ intensities needed for this search, the Large Beam Port (LBP), which has now been constructed. %The LBP is a completely unique feature among $n$ sources worldwide. 
Optimization of the cold source for NNBAR is underway via the \euro3M Horizon2020 HighNESS project \citep{Addazi:2020nlz,Santoro:2020nke}. As the ESS is expected to run at 5 MW operation after $\gtrsim 2030$, a staged program accessing the physics questions of dark sectors through sterile $n'$ searches such as $n\rightarrow n'$, $n\rightarrow n' \rightarrow n$ and $n\rightarrow n' \rightarrow \bar{n}$ has been developed, taking advantage of the existing $n$ scattering facilities at ORNL \citep{Broussard:2017yev,Broussard:2019tgw}, and continuing with an optimized experimental setup on the lower intensity fundamental physics ANNI beamline \citep{Soldner:2018ycf} as part of the HIBEAM program \citep{Addazi:2020nlz}.
%\subsection{Ultracold neutrons}

%\textit{Leah, Alexey, Valery, Albert, Anatoli}\\
Another proposed approach to the free search for $n\rightarrow\bar{n}$ utilizes a material trap for the long-term storage of ultracold neutrons. With a UCN source production of $10^8\, n$/s, the increase of the experimental sensitivity can be about 10-40$\times$ILL, and so reaching $\tau_{n\bar{n}}\sim10^{8-9}\,$s, depending on the model of $n$ reflection from the material trap walls \citep{Serebrov:2016rvi,Fomin:2017aiz,Fomin:2017lej,Fomin:2018qrq,Fomin:2019oje,Fomin:2019oyj}. The sensitivity of the experiment with UCN is lower than in the baseline NNBAR beam experiment at the ESS; however, realization of the experiment with UCN is less expensive and much more compact. In addition, this approach presents an important opportunity to perform a free search in an independent experiment using a very different methodology. 

%\subsection{Bound neutrons}
%\textit{Yeon-jae, Josh, Linyan, Ed, Georgia, Jean-Marc, Elena}\\
In similarity to free $n$ searches, observable rates for intranuclear dinucleon processes, including $n\rightarrow \bar{n}$, show great complementary experimental reach across large underground experiments such as SK \citep{Abe:2011ky,WanNeutrino2020}, DUNE \citep{Abi:2020evt}, and HK \citep{Abe:2018uyc}. 
SK has produced the world's best lower limit, $\tau_{n\bar{n}}>2.7\times10^8\,$s \citep{Abe:2011ky}. 
Prodigious amounts of $n$'s in these large mass detectors provide the capacity to overcome expected intranuclear suppression of $n\rightarrow\bar{n}$ rates \citep{Barrow:2019viz,Oosterhof:2019dlo}, though irreducible atmospheric $\nu$ backgrounds seem to persist at great cost to signal efficiency \citep{WanNeutrino2020}. Similarly, when comparing to background, intranuclear final state interactions of annihilation-generated mesons can lead to some uncertainty surrounding the region of interest when investigating reconstructed total momentum and total invariant mass \citep{Barrow:2019viz,Abe:2011ky,WanNeutrino2020}. Better modeling of the annihilation location, process, transport, and differences across many nuclear model configurations are all currently being investigated. Given the special expected topological aspects of $\bar{n}$ annihilation within nuclei, there has been much progress to date in applications of deep learning and other automated analysis techniques such as boosted decision trees to the separation of these rare process signals from background. 
%These remain burgeoning techniques, and are expected to greatly improve sensitivities in the long run. 
When converting through the traditional intranuclear suppression factor formalism \citep{Dover:1985hk,Barrow:2019viz}, intranuclear searches are expected to probe $\tau_{n\bar{n}}\gtrsim10^{8-9}\,$s.

%\subsection{Complementarity at colliders}
%\textit{David, Gustaaf, Bhupal}\\
TeV-scale colored scalars responsible for dinucleon decay, $n\rightarrow\bar{n}$, and low-scale baryogenesis can be searched for at the LHC via dijet resonances. Current LHC limits on heavy scalar diquarks are already very stringent: $M_{qq}\gtrsim 7.5$ TeV~\citep{Sirunyan:2019vgj}. This could be further improved at the future HL-LHC, and provide a complementary probe of $\Delta \mathcal{B}=2$ processes. In the context of a given model with specific flavor structures, such as PSB~\citep{Babu:2013yca}, the LHC limit could be somewhat relaxed, especially if there is a sizable branching ratio to final state quarks involving the third generation. These channels, like $tj$ and $tb$, are directly relevant for $n\rightarrow\bar{n}$ and should be searched for in future dijet analyses; such future collider constraints are expected to close portions of interesting parameter space to future free $n\rightarrow\bar{n}$ searches. A future $100\,$TeV collider could in principle probe the \textit{entire} allowed parameter space of compelling PSB models.

%We have endeavored to give an overview of many of the important contributions from many of our esteemed colleagues given at this highly successful virtualized ACFI program coordinated with Snowmass. Such contributions merit great discussion within the larger high-energy community, and show the necessity of prioritizing $\mathcal{(B-L})$-violating $\Delta \mathcal{B}=2$ experiments, including \textit{but not limited to} utilizing free cold $n$ beams at the ESS for direct $n\rightarrow\bar{n}$ searches. We look forward to the ensuing discussions to collaboratively distill new and exciting ideas from across the community throughout this dynamic process.

%In conclusion, b
By taking advantage of recent theoretical and experimental advances and next-generation facilities, % such as DUNE, Hyper-K, and the ESS, 
searches for $n\rightarrow\bar{n}$ can be performed with significantly improved sensitivity compared to previous limits, and with great complementarity to future collider-based searches. 
%Future searches will exploit improvements in modeling of intranuclear events and better signal:background discrimination, the major leaps in advanced neutronics and detection capabilities in the decades since the last free search, as well as recent advances in (theoretical approaches). 
To capitalize on these opportunities, scientific investment is needed in next decade to explore new ideas and directions which can improve the viability and sensitivity of these searches. 
$\Delta \mathcal{B} =2$ searches serve an important and complementary role to searches for neutrinoless double $\beta$-decay and proton decay, and these efforts will address an important gap in the worldwide program to understand the baryon asymmetry of the universe.

%\mychapter{2}{Acknowledgements}
%\mychapter{2}{Author Index}

\newpage
\bibliographystyle{style}    % APS-like style for physics
\bibliography{bibliography}   % name your BibTeX data base

\end{document}